\documentclass[journal,twocolumn,final]{IEEEtran}

\usepackage[pdftex]{graphicx}
\DeclareGraphicsExtensions{.pdf,.jpeg,.png}
\usepackage[cmex10]{amsmath}
\usepackage{amssymb,amsfonts,amsthm,graphicx,cite}
\usepackage{algorithmic}
\usepackage{algorithm}
\usepackage{booktabs}
\usepackage{balance}
\usepackage{subfigure}

\begin{document}

\title{Exploiting a Geometrically Sampled Grid in the SRP-PHAT for Localization Improvement and Power Response Sensitivity Analysis}

\author{Daniele~Salvati, 
				~Carlo~Drioli, 
        and~Gian~Luca~Foresti, 
\thanks{D. Salvati, C. Drioli, and G.L. Foresti are with the Department
of Mathematics and Computer Science, University of Udine, Udine 33100, Italy, e-mail: daniele.salvati@uniud.it, carlo.drioli@uniud.it, gianluca.foresti@uniud.it.}
}

\maketitle

\begin{abstract}
The steered response power phase transform (SRP-PHAT) is a beamformer method very attractive in acoustic localization applications due to its robustness in reverberant environments. This paper presents a spatial grid design procedure, called the geometrically sampled grid (GSG), which aims at computing the spatial grid by taking into account the discrete sampling of time difference of arrival (TDOA) functions and the desired spatial resolution. A new SRP-PHAT localization algorithm based on the GSG method is also introduced. The proposed method exploits the intersections of the discrete hyperboloids representing the TDOA information domain of the sensor array, and projects the whole TDOA information on the space search grid. The GSG method thus allows to design the sampled spatial grid which represents the best search grid for a given sensor array, it allows to perform a sensitivity analysis of the array and to characterize its spatial localization accuracy, and it may assist the system designer in the reconfiguration of the array. Experimental results using both simulated data and real recordings show that the localization accuracy is substantially improved both for high and for low spatial resolution, and that it is closely related to the proposed power response sensitivity measure.
\end{abstract}


\IEEEpeerreviewmaketitle

\section{Introduction}

\IEEEPARstart{T}{he} problem of locating acoustic sources is a fundamental task in applications of acoustic scene analysis and acoustic situational awareness, and it received significant attention in the research community. 
Direct methods based on the processing and fusion of data collected from microphone arrays are very attractive in acoustic applications due to their robustness and fast implementation \cite{omologo1998,DiBiase2001,Aarabi2003, Ward2003,Pertila2008,Velasco2012}. 

The steered response power phase transform (SRP-PHAT) \cite{DiBiase2001} is one of the most effective direct methods for the localization of acoustic sources in reverberant environments. It is based on a steered beamformer, which can be implemented using a space search procedure, and a map that links each position of the search grid to the time difference of arrival (TDOA) functions related to the sensor pairs. The source position is then estimated by maximization of a specific function that provides a coherent value from the entire system of microphones. The localization function is the sum of the generalized cross-correlation phase transform (GCC-PHAT) \cite{Knapp1976} values estimated from all combinations of microphone pairs. 

The use of an acoustic map related to the TDOA between two microphones has been first introduced in 1998 by Omologo and De Mori \cite{omologo1998}. The authors 
call this procedure global coherence field (GCF), and introduce the GCF-PHAT \cite{Brutti2010} method, which is equivalent to SRP-PHAT. In 2001, the authors of \cite{DiBiase2001} demonstrated that the SRP-PHAT can be computed by decomposing
the steered beamformer into the sum of the beamformers corresponding to the sensor pairs of the array, and that the steered
response of two sensors is equivalent to the GCC-PHAT function. Thus, the SRP-PHAT is effectively computed by using the GCF and the GCC-PHAT, making its practical implementation very attractive. In fact, the GCC-PHAT can be computed in the frequency domain using
the fast Fourier transform (FFT) for each sensor pair, and the acoustic map can be computed by access and sum operations on a look-up table of GCC-PHAT values. The sampled space grid, which is a set of candidate positions for the source, is pre-calculated defining a look-up table that links the position in space with TDOA values of microphone pairs. 

Note that the SRP-PHAT algorithm is actually the combination of two distinct components: the steered response power (SRP) computation and a PHAT prefiltering. The role of the PHAT filter is to normalize the narrowband steered beamformer and to only take into account the phases of the cross-power spectral density. The normalization has the positive effect of increasing the spatial resolution \cite{Salvati2014}, and it is one of the advantage of this method in a reverberant environment since it allows improved identification of direct paths and reflections.

Most part of the past researches on SRP-PHAT focused on solutions to reduce the computational cost of the grid-search step. In some cases, the problem has been faced by calculating the steered response on a limited set of candidate source positions, e.g, by using a stochastic region contraction \cite{Berger1991}, by using a generic doubly hierarchical search algorithm \cite{Zotkin2004}, or by only considering the larger GCC-PHAT coefficients \cite{Dmochowski2007}. However, these methods usually discard part of the information available and the localization performance can degrade when reverberation increases \cite{Nunes2014}. In \cite{Dmochowski2007}, since the GCC-PHAT function provides different local maxima due to the contribution of direct-path and early reflections, when the direct-path peak has lower intensity with respect to a reflection peak, the peak picking procedure returns a wrong contribution since it disregards the direct-path peak in favor of a reflection peak. 

Recently, a method that relies on the use of a coarser grid has been proposed in \cite{Cobos2011}. Herein it is shown that the traditional grid-search approach of SRP-PHAT degrades its performance when the spatial resolution decreases due to the loss of information of GCC-PHAT functions. To face this problem, in \cite{Cobos2011} a scalable spatial sampling (SSS) is proposed to accumulate the GCC-PHAT values in a range that covers the volume surrounding each point of the defined spatial grid. The GCC-PHAT accumulation limits are determined by the gradient of the inter-microphone time delay function corresponding to each microphone pair. The reduced number of spatial grid points involves a lower computational cost, but the accuracy is limited by the resolution of the grid. Other methods have been proposed that improve the localization accuracy by refining the search procedure from a coarser grid to a finer grid using iterative searching procedures \cite{Marti2013,Nunes2014,Lima2015}.

The above mentioned methods have in common the way in which the space search grid is designed, and the way in which the relationship between the points on the grid and the TDOAs of microphone pairs is build. Specifically, for each microphone pair and for each point on the grid, an unique integer TDOA value is selected to be the acoustic delay information linked to that point. This uniform regular grid (URG) procedure does not guarantee that all TDOA samples are associated to points on the grid, nor that the spatial grid is consistent since some of the points in the grid may not correspond to an intersection of a bare minimum of three hyperboloids (or two hyperbolas, in 2D). The linking from space points on the grid to TDOAs also does not allow for spatial resolution scalability, since when the number of points is reduced, part of the TDOA information gets lost as it results no more associated to any points on the grid. For these reasons, different methods have been proposed in \cite{Cobos2011,Marti2013,Nunes2014} to collect and use the TDOA information related to the volume surrounding each spatial point on the search grid. A boundary-vertex (BV) approach is used in \cite{Nunes2014}, in which the GCC-PHAT accumulation limits are determined by the cube surrounding the volume vertices. In \cite{Marti2013}, a modified SSS (MSSS) is proposed , which exploits the mean of the accumulated GGC-PHAT values for each volume. However, these methods does not take into account how TDOA information is distributed in the space. We will see that the spatial distribution of all TDOA information is an important information that can be used to compute a sensitivity measure of the acoustic system with respect to the search region and to improve the localization accuracy. 
There is thus the need of a rigorous analysis of the spatial grid map and of how the TDOA information from GCC-PHAT functions is accumulated in the space. 

In this paper, we study the properties of the SRP-PHAT algorithm focusing especially on the grid resolution, which is in general arbitrarily imposed depending on the type of application, and the TDOA resolution, which is given by the distance between the microphones and the sample rate used in the digital system. We propose a new spatial grid design procedure, named geometrically sampled grid (GSG), which makes use of the discrete hyperboloids (representing all possible locations related to a TDOA) and of their intersections, to design an acoustically-coherent space grid on which the source search can be performed.

Moreover, we will show how, based on the density analysis of hyperboloid intersections, a steered power response sensitivity analysis of the localization system can be conducted. We refer herein to \textit{sensitivity} as a quantified measure of the change of the  response power with respect to the change of the spatial position, predicting where the search space will be characterized by higher and lower localization accuracy.
To date, studies concerning  the information distribution of SRP-like localization methods are not frequent in the literature. An example is \cite{Nunes2012}, in which a discriminability measure is proposed, which only considers the array geometry and the sampling frequency to distinguish a given point in space from its neighbors.  
In contrast with it, the proposed GSG includes in the analysis process a relationship between the sampled space and all discrete samples of the GCC-PHAT functions to prevent the loss of information that may arise from the choice of an arbitrary desired spatial resolution. 

Besides that, the coherent sample grid and the power response sensitivity analysis are useful tools to decide if the spatial resolution and the sensitivity map of a given array configuration are adequate and, if not, to assist the system designer in its reconfiguration (e.g., by the positioning of additional sensors or by increasing the sampling frequency). Hence, it means that the system configuration designed by the GSG procedure generates a grid in which each point is consistent for the localization, i.e. it is the point of intersection of at least three hyperboloids.

With respect to other approaches whose aim is to improve the localization accuracy, the GSG method builds the steered power response function using all the TDOA information available from the GCC-PHAT functions related to the sensor pairs in the array, it solves the problem of arbitrarily selecting the spatial grid resolution without loss of information, and it turns out to notably improve the localization performances. The geometric approach based on the analysis of hyperboloid intersections allows the design of a sensitivity map, in which the regions where the localization is more accurate correspond to the high sensitivity regions of the steered power response function.

Finally, the GSG method might also provides reduced computational cost with respect to the URG method in three cases: 1. when the search procedure is restricted to the coherent grid, thus discarding the URG points which are not covered by sufficient acoustic information, 2. when the type of application allows to use a coarser grid and a lower spatial resolution, 3. when the search can be restricted only to the high sensitivity regions, in which the localization accuracy is maximized.  

The paper is organized as follows. After presenting the relationship between the spatial grid and the TDOA functions in Section \ref{sec:sm}, the SRP-PHAT method is described in Section \ref{sec:srp}. In Section \ref{sec:gsg} the GSG algorithm and the GSG based SRP-PHAT are presented. Finally, Section \ref{sec:er} illustrates experimental results obtained in a simulated reverberant environment and in a real-world scenario.

\section{Spatial Grid and Time Difference of Arrival}
\label{sec:sm}

Consider a reverberant room, and a location volume $G=(G_x \times G_y \times G_z)$, discretized with a space resolution $\Delta$, in which the acoustic source is searched. A generic grid  position is denoted by $\mathbf{r}_g=[x_g \quad y_g \quad z_g]^T$, $\mathbf{r}_g \in G$. Within the room, we suppose $M$ microphones disposed according to a given geometry. The positions of the $M$ microphones in Cartesian coordinates are
\begin{equation}
\mathbf{r}_m = [x_m \quad y_m \quad z_m]^T, \quad m=1,2,\dots,M
\end{equation}
where $(\cdot)^T$ denotes the transpose operator. We will consider all possible sensor pairs of the array in our analysis.  Accordingly, an array of $M$ microphones provides $N$ unique microphone pairs, with
\begin{equation}
N = \binom{M}{2}.
\end{equation}
Given a generic sensor pair $n$, referred to two microphones located in $\mathbf{r}_i$ and $\mathbf{r}_j$, the maximum TDOA in samples $T_n \in \mathbb{Z}$ is obtained as
\begin{equation}
T_n = \text{fix}\Bigl(\frac{||\mathbf{r}_i-\mathbf{r}_j|| f_s}{c}\Bigl)
\end{equation}
where fix$(\cdot)$ denotes the round toward zero operation, $f_s$ is the sampling frequency, $c$ is the speed of sound, and $||\cdot||$ denotes Euclidean norm.
The admissible range of values for the TDOA is [-$T_n$,$T_n$], thus the possible TDOA values for the sensor pair $n$ are $2 T_n+1$. 

We study the case in which a single acoustic source is active at time $k$ and the unknown coordinate position is
\begin{equation}
\mathbf{r}_s(k) = [x_s(k) \quad y_s(k) \quad z_s(k)]^T.
\end{equation}
The observed signals are given by the convolution of the unknown source $s(k)$ with corresponding acoustic impulse responses $h_m$ from the source to the microphone $m$. The reverberant model for discrete-time signals can be expressed as
\begin{equation}
\tilde{x}_m(k)=h_m \ast s(k)+v_m(k)
\end{equation}
where $m=1,2,\dots,M$, $\ast$ denotes convolution, $v_m(k)$ is the uncorrelated noise signal.
The relationship between a generic space position $\mathbf{r}_g$ and the TDOA of the wavefront at the sensor pair $n$ of two microphones $i$ and $j$ becomes 
\begin{equation}
\tau_{n}(\mathbf{r}_g)=\text{round}\Bigl[\frac{(||\mathbf{r}_g-\mathbf{r}_i||-||\mathbf{r}_g-\mathbf{r}_j||) f_s}{c}\Bigl]
\label{hyp}
\end{equation}
where round$[\cdot]$ denotes rounding operator. Note that equation (\ref{hyp}) assumes that the TDOA is an integer and it is expressed in samples.
Equation (\ref{hyp}) represents an hyperboloid, which describes the locus of possible sound source locations generating the same TDOA for that microphone pair. To uniquely determine the position of the source (the three unknown coordinates), we need, at a bare minimum, a system of three equations providing the intersection of the three hyperboloids.

The spatial grid in the SRP-PHAT algorithm is traditionally calculated with an URG approach that links the uniformly distributed points on the spatial grid to TDOAs related to the sensor pairs. 

Given a look-up table $\chi(\mathbf{r}_g,n)$ which stores the relationship between grid positions and TDOAs, the URG procedure is summarized in Algorithm \ref{algurg}. The limitations of this approach are that it does not guarantee that all TDOA values correspond to a point on the space grid (and if this is the case, the information related to that TDOA is lost),  and that it is not guaranteed that every point of the grid is consistent with the condition of being the locus where at least three hyperboloids intersect. Note that, due to the rounding operator, from the URG point of view everything goes as if in each grid position there is an intersection of $N$ hyperboloids. The approximation due to the rounding operation can link a whole set of neighbor points to the same TDOA, resulting in practice in an uniform steered response power in that region.

\begin{algorithm}[t]
\caption{URG Algorithm}
\label{algurg}
\begin{algorithmic}
\small
\STATE $N$: number of microphone pairs  
\FOR{all $\mathbf{r}_g \in G$}
	\FOR{$n=1$ to $N$}
		\STATE Calculate $\tau_{n}(\mathbf{r}_g)$ by means of Eq. (\ref{hyp})
		\STATE $\chi(\mathbf{r}_g,n)=\tau_{n}(\mathbf{r}_g)$
	\ENDFOR
\ENDFOR
\end{algorithmic}
\end{algorithm}

\section{Steered Response Power Phase Transform}
\label{sec:srp}

The steered beamformer for source localization is based on the computation of a filtered combination of the delayed signals sensed by the array.   
Typically, a broadband steered power beamformer is computed in the frequency-domain by applying a FFT on a portion of the signal and by calculating the response power on each frequency bin. Subsequently, a fusion of these estimates is computed. 
The narrowband output signal of a delay and sum beamforming can be expressed as 
\begin{equation}
Y(f,\mathbf{r}_g,k)=\mathbf{A}^H(f,\mathbf{r}_g)\mathbf{X}(f,k)
\end{equation}
where $f$ is the frequency index, the superscript $H$ represents the Hermitian (complex conjugate) transpose, $\mathbf{A}(f,\mathbf{r}_g)$ is the steering vector corresponding to a given position $\mathbf{r}_g$, $\mathbf{X}(f,k)=[X_1(f,k) \: X_2(f,k) \dots X_M(f,k)]^T$,  $Y(f,\mathbf{r}_g,k)$ and $X_m(f,k)$, $m=1,2,\dots,M$, are the FFT of the signals.
A formal way to express the SRP-PHAT using the beamforming notation in time-frequency domain with an incoherent arithmetic mean is given by 
\begin{equation}
\begin{split}
 P(\mathbf{r}_g,k)&=\sum_{f=0}^{L-1} E\{|Y(f,\mathbf{r}_g,k)|^2\}\\
&=\sum_{f=0}^{L-1} \mathbf{A}^H(f,\mathbf{r}_g)(\mathbf{\Phi}(f,k) \div |\mathbf{\Phi}(f,k)|)\mathbf{A}(f,\mathbf{r}_g)
\end{split}
\label{srp}
\end{equation}
where $P(\mathbf{r}_g,k)$ is the power spectral density of the beamformer output at time $k$ in position $\mathbf{r}_g$, $L$ is the length of the FFT analysis window,  $E\{\cdot\}$ denotes mathematical expectation, $\mathbf{\Phi}(f,k)$ is the cross-spectral density matrix, $\div$ denotes element-wise division, and $|\cdot|$ denotes element-wise absolute value operation. The PHAT filter discards the magnitude and only keeps the phase of $\mathbf{\Phi}(f,k)$ for computing the steered responses.

In \cite{DiBiase2001}, the authors demonstrate that SRP-PHAT can be computed by decomposing the steered beamformer as a sum of element pairs beamformers. Moreover, the steered beamformer of a two-element array is equivalent to the GCC-PHAT of those two microphones. The GCC-PHAT is estimated using the discrete Fourier transform (DFT) and the inverse DFT (IDFT), which can be efficiently implemented with the FFT, while the equation (\ref{srp}) requires the calculation of the steered beamformer for each frequency bin.

The steered response power with the URG can now be expressed as an operation of GCC-PHAT functions
\begin{equation}
 P_\text{URG}(\mathbf{r}_g,k)=\sum_{n=1}^{N} R_n[\tau_n(\mathbf{r}_g),k]
\label{srpgcf}
\end{equation}
where the GCC using the PHAT whitening for a generic $n$ pair is given by
\begin{equation}
R_n[\tau_n(\mathbf{r}_g),k]=\frac{1}{L}\sum_{f=0}^{L-1} \Psi(f,k) [X_i(f,k)X_j^*(f,k)] e^{\frac{j2\pi f \tau_n(\mathbf{r}_g)}{L}}\\
\label{gcc}
\end{equation}
in which $(\cdot)^*$ denotes the complex conjugate, and the PHAT filter is
\begin{equation}
\Psi(f,k)=\frac{1}{|X_i(f,k)X_j^*(f,k)|}.
\label{paht}
\end{equation}

The SRP-PHAT method finally estimates the source position by picking the maximum value of the power output on every point $\mathbf{r}_g$ of the search grid 
\begin{equation}
\widehat{\mathbf{r}}_s(k)= \underset{\mathbf{r}_g}{\operatorname{argmax}}[ P_\text{URG}(\mathbf{r}_g,k)].
\end{equation}
The SRP-PHAT-URG is summarized in Algorithm \ref{algsrpurg}.

\begin{algorithm}[t]
\caption{SRP-PHAT-URG}
\label{algsrpurg}
\begin{algorithmic}
\small
\STATE Initialization: for all grid position $ \mathbf{r}_g \in G$, $P_\text{URG}(\mathbf{r}_g,k)=0$
\FOR{all $\mathbf{r}_g \in G$}
	\FOR{$n=1$ to $N$}
		\STATE $P_\text{URG}(\mathbf{r}_g,k)=P_\text{URG}(\mathbf{r}_g,k)+R_{n}[\chi(\mathbf{r}_g,n),k]$
	\ENDFOR
\ENDFOR
\STATE $\widehat{\mathbf{r}}_s(k)=\underset{\mathbf{r}_g}{\operatorname{argmax}}[P_\text{URG}(\mathbf{r}_g,k)] \quad \mathbf{r}_g \in G$
\end{algorithmic}
\end{algorithm}

\section{Geometrically Sampled Grid Algorithm}
\label{sec:gsg}

The geometrically sampled grid (GSG) algorithm is based on computing the space grid map by considering the discretization of hyperboloids with a desired spatial resolution, and by taking into account all discrete TDOA values.

Consider a generic microphone pair $n$, we can interpret the equation (\ref{hyp}) as the quadratic surface of an hyperboloid in a local Cartesian system $(x_n,y_n,z_n)$ with the origin in the midpoint of the segment joining the two microphones $i$ and $j$
\begin{equation}
\frac{x^2_n}{a_1^2}-\frac{y^2_n}{a_2^2}-\frac{z^2_n}{a_3^2}-1=0
\label{hyp2s}
\end{equation}
where $a_1>0$, $a_2>0$, and $a_3>0$.
This is the equation of an hyperboloid of two sheets assuming that the $x_n$ axes is coincident with the line joining the two microphones. The transformation between the two coordinate systems $(x,y,z)$ and $(x_n,y_n,z_n)$ is computed with an operation of translation and rotation and it is expressed by
\begin{equation}
\begin{bmatrix}
x_n\\
y_n\\
z_n
\end{bmatrix}=\mathbf{\Omega}_n\mathbf{R}_n
\begin{bmatrix}
x\\
y\\
z
\end{bmatrix}
\end{equation}
where $\mathbf{\Omega}_n$ and $\mathbf{R}_n$ are respectively the translation matrix and the rotation matrix for pair $n$. Equation (\ref{hyp2s}) can be decomposed in a simpler form as an hyperbola that is rotated along  the $x_n$ axis. By including the information in $\tau_n$ for the sheet identification, the hyperbola on axes $(x_n,y_n)$ can be written in the following way
\begin{equation}
x_n=f_x(y_n)=\text{sign}(\tau_n)\sqrt{\Bigl(\frac{{y}^2_n}{a_2^2}+1\Bigl)a_1^2} 
\label{hyp2d}
\end{equation}
where sign$(\cdot)$ denotes the signum function to identify the sheet given by TDOA $\tau_n$. 
Comparing the equation (\ref{hyp}) (at $z=0$) and (\ref{hyp2d}) we have
\begin{equation}
\begin{split}
a_1&=\frac{c\tau_n}{2f_s},\\
a_2&=\sqrt{\Bigl(\frac{||\mathbf{r}_i-\mathbf{r}_j||}{2}\Bigl)^2-a_1^2}.
\end{split}
\end{equation}
If $\tilde{G}_x=\mathbf{\Omega}_n\mathbf{R}_nG_x$, $\tilde{G}_y=\mathbf{\Omega}_n\mathbf{R}_nG_y$, and $\tilde{G}_z=\mathbf{\Omega}_n\mathbf{R}_nG_z$, we call 
\begin{equation}
y_n^r=i\Delta, \quad i \in [i^y_\text{min},i^y_\text{max}]
\end{equation}
the discretization of $\tilde{G}_y$ with resolution step $\Delta$, and we can calculate the grid points $x_n \in \tilde{G}_x$  from (\ref{hyp2d}) and its discrete values as 
\begin{equation}
x_n'=\text{round}\Bigl[\frac{f_x(y_n^r)}{\Delta}\Bigl] \Delta.
\end{equation}
We can now consider the circumference of radius $y_n^r$ for estimating the rotation of the hyperbola along the $x_n$ axes. Then, we have for all $z'_n \in \tilde{G}_z$
\begin{equation}
\begin{split}
z'_n&=i \Delta, \quad i \in [i^z_\text{min},i^z_\text{max}],\\
y'_n&=\pm\text{round}\Bigl[\frac{\sqrt{(y_n^r)^2-(z_n')^2}}{\Delta}\Bigl] \Delta, \quad y_n' \in \tilde{G}_y.
\end{split}
\end{equation}
With this procedure the $\Delta$ spatial resolution is guaranteed for the y-axis and the z-axis, but not for
the x-axis. We can then rewrite equation (\ref{hyp2d}) in the following form
\begin{equation}
y_n=f_y(x_n)=\text{sign}(\tau_n)\sqrt{\Bigl(\frac{{x}^2_n}{a_1^2}-1\Bigl)a_2^2}.
\label{hyp2xd}
\end{equation}
We now call
\begin{equation}
x_n''=i\Delta, \quad i \in [i^x_\text{min},i^x_\text{max}]
\end{equation}
the discretization of $\tilde{G}_x$ with resolution step $\Delta$. We can now calculate the grid points $y_n \in \tilde{G}_y$  from (\ref{hyp2xd}) and their discrete values 
\begin{equation}
y_n^r=\text{round}\Bigl[\frac{f_y(x_n^r)}{\Delta}\Bigl] \Delta.
\end{equation}
If $(x_n'',y_n^r)\neq(x_n',y_n^r)$, a new grid point is calculated, and the circumference of radius $y_n^r$ in $x_n''$ can be considered for estimating the rotation of the hyperbola along the $x_n$ axes, obtaining the coordinates $(x_n'', y_n'', z_n'')$. This procedure ensures that also the x-axis will eventually have spatial resolution $\Delta$.

After the transformation of $\mathbf{r}_n=[x_n' \quad y_n' \quad z_n']^T$ (or $\mathbf{r}_n'=[x_n'' \quad y_n'' \quad z_n'']^T$) into the coordinate system $(x,y,z)$, we obtain the grid sample position $\mathbf{r}_g=[x_g \quad y_g \quad z_g]^T$. Note that, due to the rounding operator, there are regions where two or more hyperboloids corresponding to different TDOAs may be mapped on the same point of the grid. Thus, in contrast to the URG case in which, due to equation (\ref{hyp}), there are always exactly $N$ TDOA values associated to each point on the grid (one for each microphone pair), the GSG procedure may associate less than $N$, $N$ or more than $N$ TDOAs to a point on the grid. This property is illustrated in Figure \ref{num_ype}, for a section of the search space corresponding to a simulated acoustic environment.      
\begin{figure}[!t]
\centering
\includegraphics[width=3.4in]{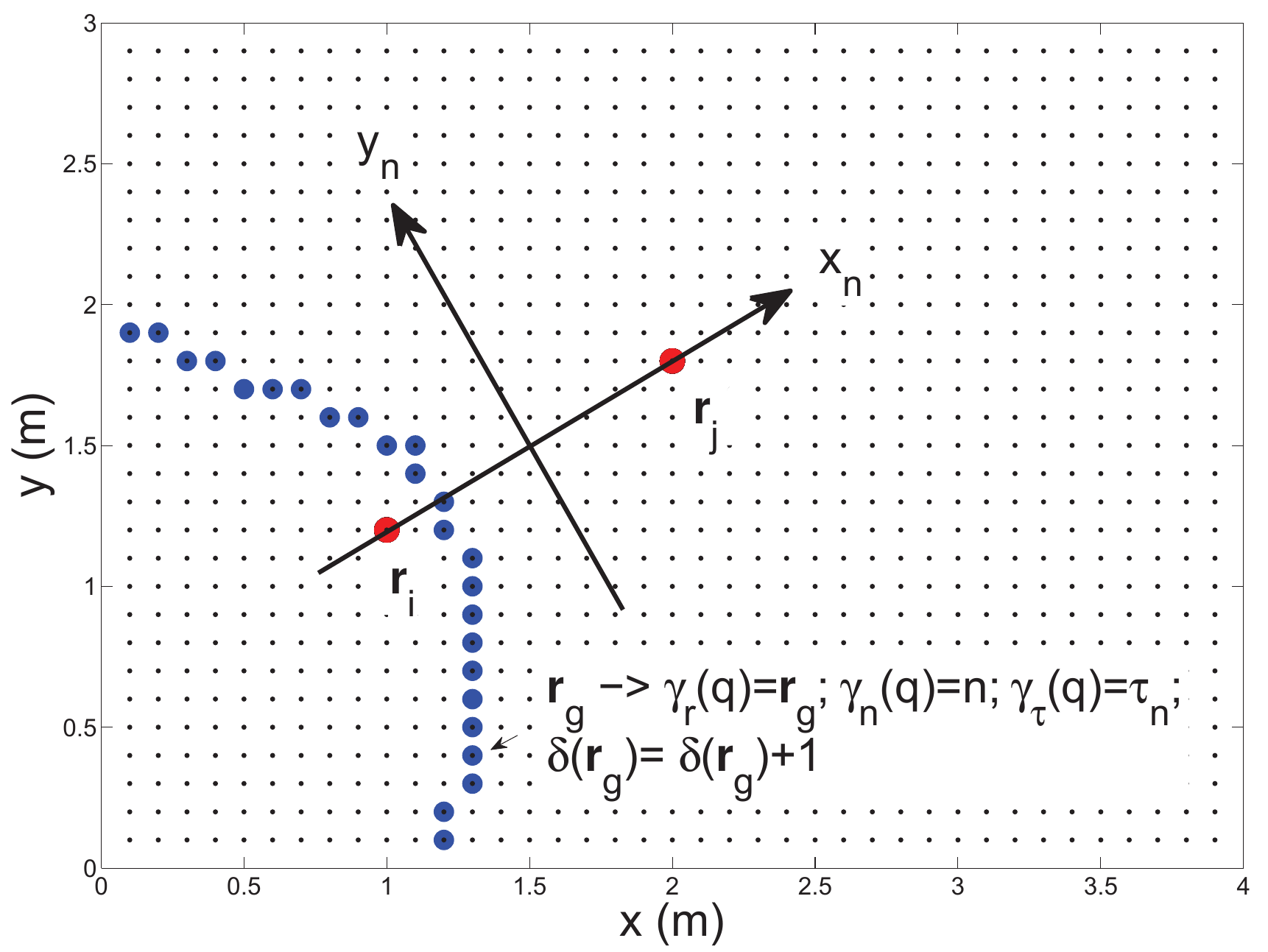}
\caption{A discrete hyperbola related to a TDOA $\tau_n=-90$ samples using the GSG algorithm for a microphone pair $\mathbf{r}_i=[1 \quad 1.2]^T$ m and $\mathbf{r}_j=[2 \quad 1.8]^T$ m. For each grid sample position $\mathbf{r}_g$ of the hyperbola, the values $\mathbf{r}_g$, $n$, and $\tau_n$ are stored in look-up tables $\gamma_r(q)$, $\gamma_n(q)$ and $\gamma_{\tau}(q)$ respectively, and the number of hyperbolas passing through position $\mathbf{r}_g$ are stored in $\delta(\mathbf{r}_g)$. Space resolution $\Delta$ is 0.1 m and $f_s=44.1$ kHz.}
\label{xymap}
\end{figure}

We build the grid map with resolution $\Delta$ for all $N$ microphone pairs and for each pair considering all $2 T_n+1$ TDOA values. The values of the discrete hyperboloid and the TDOA information are stored in four look-up tables. 
To each discrete hyperboloid point, we assign an index $q$, so that we have a table $\gamma_r(q)$ for the position, a table $\gamma_n(q)$ for the pair index, and a table $\gamma_{\tau}(q)$ for the TDOA. The tables are used in real-time for estimating the acoustic energy and computing the accumulation of GCC-PHAT functions by all considered sensor pair. We define $Q'$ as the number of discrete hyperboloid points calculated by the GSG algorithm.
The last look-up table, which we name $\delta(\mathbf{r}_g)$, contains the actual number of the surfaces intersecting in position $\mathbf{r}_g$. 

To be consistent with the definition of a candidate source position as the intersection of hyperboloids, the following constraint is applied after the complete analysis of $\delta(\mathbf{r}_g)$ for all $\mathbf{r}_g \in G$
\begin{equation}
\delta(\mathbf{r}_g)=0, \quad \text{if} \quad \delta(\mathbf{r}_g) < \mu 
\label{acon}
\end{equation}
where $\mu=3$ and $\mu=2$ in case of 3D and 2D localization respectively. The constraint has the goal to discard those sample space point that are not consistent for the localization. The inconsistent grid points are eliminated from the look-up tables $\gamma_r(q)$, $\gamma_n(q)$, and $\gamma_{\tau}(q)$ so that all information on the coherent grid representing the relationship with TDOAs of all pair sensor can be used for the localization. If $T$ is the number of points which are non consistent with respect to condition (\ref{acon}), then $Q=Q'-T$ is the number of discrete hyperboloid points after their removal. Figure \ref{xymap} shows a discrete hyperbola related to a TDOA $t_n=-90$ samples of a specific microphone pair $n$. The space resolution is $\Delta=0.1$ m, and the area of analysis is $G_x=4$ m and $G_y=3$ m. Blue circles are the identified grid positions that are stored in the look-up tables $\gamma_r(q)$, $\gamma_n(q)$, $\gamma_{\tau}(q)$ and $\delta(\mathbf{r}_g)$. The table $\delta(\mathbf{r}_g)$ is the sensitivity map that gives information on how all sampled GCC-PHAT values are projected into space. In this way, we can obtain a sensitivity map of the considered grid. It will be shown in the experimental section that an improvement in the localization accuracy is obtained in the high sensitivity regions, where the accumulation of GCC-PHAT information is higher.
The coherent grid $\Gamma_r$ related to the array is calculated by removing duplicate positions in $\gamma_r(q)$
\begin{equation}
\Gamma_r = \text{unique}[\gamma_r(q)]
\end{equation}
where unique$(\cdot)$ denotes the operator which removes duplicate values from a list.

The procedure to build the coherently sampled grid and the sensitivity map in a geometric way is given by the following steps:
\begin{enumerate}
\item Initialization of $\delta(\mathbf{r}_g)=0$ for all $\mathbf{r}_g \in G$ and of index q=0;
\item For each sensor pair $n=1,2,\dots,N$ and for all TDOA values $\tau_n$ in the range [-$T_n$,$T_n$], calculate the discrete hyperboloid, write the values in the look-up tables $\gamma_r(q)$, $\gamma_n(q)$, and $\gamma_{\tau}(q)$, update the value of the look-up table $\delta(\mathbf{r}_g)=\delta(\mathbf{r}_g)+1$, and update $q=q+1$;
\item After the geometric discrete analysis of hyperboloids has terminated, apply the constraint on $\delta(\mathbf{r}_g)$ and update the look-up tables $\gamma_r(q)$, $\gamma_n(q)$, and $\gamma_{\tau}(q)$.
\end{enumerate}
The GSC algorithm is summarized in Algorithm \ref{alg1}.

\begin{algorithm}[t]
\caption{GSG Algorithm}
\label{alg1}
\begin{algorithmic}
\small
\STATE $N$: number of microphone pairs  
\STATE $\Delta$: spatial resolution  
\STATE Initialization: for all grid position $\mathbf{r}_g \in G$, $\delta(\mathbf{r}_g)=0$
\STATE Initialization: $q=0$
\FOR{$n=1$ to $N$}
	\STATE Calculate the local coordinate system ($x_n,y_n,z_n$)
	\STATE Calculate $2 T_n+1$ (number of TDOA samples for the $n$th pair)
	\FOR{$\tau_n=-T_n$ to $T_n$}
		\FOR{all $y_n^r \in \tilde{G}_y$}
			\STATE Calculate $x_n'$
			\IF{$x_n' \in \tilde{G}_x$}
				\FOR{all $z_n' \in \tilde{G}_z$}
					\STATE Calculate $y_n'$
					\IF{$y_n' \in \tilde{G}_y$}
						\STATE Transform $\mathbf{r}_n=[x_n'  \; y_n' \; z_n']^T$ to $\mathbf{r}_g=[x_g  \; y_g  \; z_g]^T$
						\STATE $\gamma_r(q)=\mathbf{r}_g$, $\gamma_n(q)=n$, $\gamma_{\tau}(q)=\tau_n$
						\STATE $\delta(\mathbf{r}_g)=\delta(\mathbf{r}_g)+1$
						\STATE q=q+1
					\ENDIF
				\ENDFOR
			\ENDIF
		\ENDFOR
		\FOR{all $x_n'' \in \tilde{G}_x$}
			\STATE Calculate $y_n^r$
			\IF{$y_n^r \in \tilde{G}_x$ and $(x_n'',y_n^r)\neq(x_n',y_n^r)$}
			\FOR{all $z_n'' \in \tilde{G}_z$}
					\STATE Calculate $y_n''$
					\IF{$y_n'' \in \tilde{G}_y$}
						\STATE Transform $\mathbf{r}'_n=[x_n''  \; y_n'' \; z_n'']^T$ to $\mathbf{r}'_g=[x'_g  \; y'_g  \; z'_g]^T$
						\STATE $\gamma_r(q)=\mathbf{r}'_g$, $\gamma_n(q)=n$, $\gamma_{\tau}(q)=\tau_n$
						\STATE $\delta(\mathbf{r}'_g)=\delta(\mathbf{r}'_g)+1$
						\STATE q=q+1
					\ENDIF
			 \ENDFOR
			\ENDIF
		\ENDFOR
	\ENDFOR
\ENDFOR
\STATE Q'=q
\STATE Apply the constraint and compute $T$
\STATE Update $\gamma_r(q)$, $\gamma_n(q)$, and $\gamma_{\tau}(q)$
\STATE Q=Q'-T
\STATE $\Gamma_r = \text{unique}[\gamma_r(q)]$
\end{algorithmic}
\end{algorithm}

\begin{algorithm}[t]
\caption{SRP-PHAT-GSG}
\label{alg2}
\begin{algorithmic}
\small
\STATE Initialization: for all grid position $ \mathbf{r}_g \in \Gamma_r$, $P_\text{GSG}[\mathbf{r}_g,k]=0$
\FOR{$q=1$ to $Q$}
	\STATE  $P_\text{GSG}[\gamma_r(q),k]=P_\text{GSG}[\gamma_r(q),k]+R_{\gamma_n(q)}[\gamma_{\tau}(q),k]$
\ENDFOR
\STATE $\widehat{\mathbf{r}}_s(k)=\underset{\mathbf{r}_g}{\operatorname{argmax}}(P_\text{GSG}[\mathbf{r}_g,k]) \quad \mathbf{r}_g \in \Gamma_r$
\end{algorithmic}
\end{algorithm}

Finally, at each analysis frame $k$, the GSG based SRP-PHAT is computed in three steps. First, the map is initialized by imposing the steered response power $P_\text{GSG}[\mathbf{r}_g,k]=0$ with $\mathbf{r}_g \in \Gamma_r$. Then, the values from the estimated GCC-PHAT functions are accumulated in the grid map. Finally, the source position is estimated by picking the maximum value of the acoustic map. The SRP-PHAT-GSG is summarized in Algorithm \ref{alg2}.

The output of the SRP-PHAT using the GSG algorithm can be expressed as
\begin{equation}
P_\text{GSG}(\mathbf{r}_g,k)=\sum_{h \in H_r} R_{\gamma_n(h)}[\gamma_\tau(h),k]
\label{srpgsgp}
\end{equation}
where 
\begin{equation}
H_r=\{i:\gamma_r(i)=\mathbf{r}_g\}
\end{equation}
are the look-up table indices corresponding to the TDOAs for the position $\mathbf{r}_g \in \Gamma_r$ of all the $N$ sensor pairs. Note that $H_r$ is a set of TDOAs of dimension $\delta(\mathbf{r}_g)$. After some manipulation on equation (\ref{srpgsgp}), we can write the SRP-PHAT-GSG as
\begin{equation}
P_\text{GSG}(\mathbf{r}_g,k)=\sum_{n=1}^{N}\sum_{z \in Z_{r,n}} R_{n}[\gamma_\tau(z),k]
\label{srpgsg}
\end{equation}
where 
\begin{equation}
Z_{r,n}=\{i:[\gamma_r(i)=\mathbf{r}_g]\land[\gamma_n(i)=n]\}
\end{equation}
are the look-up table indices corresponding to the TDOAs for the position $\mathbf{r}_g \in \Gamma_r$ of the sensor pair $n$. Note that $Z_{r,n}$ is an empty set if $\{i:[\gamma_r(i)=\mathbf{r}_g]\land[\gamma_n(i)=n]\}$ is null.
By comparing equations (\ref{srpgcf}) and (\ref{srpgsg}), we can observe that for each position related to the microphone pair $n$, we can have a larger amount of TDOA information, which is the principal reason of the increased localization performance in the high sensitivity region. Note that the SRP-PHAT expressed by equation (\ref{srpgsg}) has a similar form of other accumulation methods \cite{Cobos2011,Marti2013,Nunes2014}. However, GSG designs a coherent spatial grid and provides a sensitivity map, which gives information of how the whole GCC-PHAT information is distributed in the search space, resulting in different regions characterized by different localization accuracies. 

The computational cost for the GSG algorithm is equivalent to that of the URG procedure for computing the power map, since for both algorithms the relationship between TDOAs and positions in space is pre-calculated offline using the look-up tables, and online summation is negligible. Consistent reduction of the computational cost may occur for the search procedure, which depends on the number of sample grid positions. If the search procedure is restricted to the coherent grid, the computational cost is inferior to the URG method due to the discarded points. Moreover, the computational cost may be also reduced by using a coarser grid or by only searching in the high sensitivity regions, in which the localization accuracy is maximized.   

\section{Experimental Results}
\label{sec:er}

\subsection{Spatial Grid and Power Response Sensitivity Analysis}

In this section, we present experimental results concerning the construction of the spatial grid and the analysis of the power response sensitivity using the GSG algorithm for an uniform linear array (ULA). Spatial grids were designed using different small-array sizes, sampling rate values, and spatial resolutions. A search region of 2 m $\times$ 2 m was considered. Table \ref{tr} shows the resulting number of grid points when using the URG and the GSG methods, for an ULA with an inter-microphone distance of 0.15 m. The coverage percentage values reported show how the acoustically coherent grid is in some cases much smaller if compared to the uniform regular grid (especially when using a small array size combined with a high spatial resolution). As already noted, using the coherent spatial grid obtained by the GSG algorithm in those cases, has the advantage of providing a position search domain which is consistent with the hyperboloid intersections, whereas URG grid would also contain non-consistent regions which would provide misleading information, since the corresponding energy on the search map is usually comparable to that of consistent regions.

\begin{table*}[!t]
\centering
\renewcommand{\arraystretch}{1.1}
\caption{Comparison of number of grid points for a ULA using URG and GSG algorithm.}
\label{tr}
\centering
\begin{tabular}{@{}l|l|c|c|c|c|c@{}}
\toprule
\multicolumn{2}{l|}{} & URG (M=3,4,5,6) & GSG (M=3) & GSG (M=4) & GSG (M=5) & GSG (M=6)\\ 
\midrule
$f_s$=16000 Hz & $\Delta=0.01$ m & 40000 (100 \%) & 486 (1.22 \%) & 3930 (9.83 \%) & 10854 (27.14 \%)& 20242 (50.61 \%)\\ 
 & $\Delta=0.05$ m& 1600 (100 \%) & 264 (16.50 \%) & 1140	(71.25 \%) & 1446	(90.38 \%) & 1509 (94.31 \%) \\ 
 & $\Delta=0.1$ m& 400 (100 \%) & 185 (46.25 \%)& 358 (89.50 \%) & 370 (92.50 \%)& 374 (93.50 \%) \\ 
\midrule
$f_s$=44100 Hz & $\Delta=0.01$ m & 40000 (100 \%) & 3710 (9.28	 \%) & 15816 (39.54 \%) & 29708 (74.27 \%) & 36958 (92.40 \%) \\ 
 & $\Delta=0.05$ m& 1600 (100 \%) & 1281 (80.06 \%) & 1527 (95.44 \%) & 1540 (96.25 \%) & 1559 (97.44 \%)\\ 
 & $\Delta=0.1$ m& 400 (100 \%) & 372 (93.00 \%) & 378 (94.50 \%) & 380 (95.00 \%) & 380 (95.00 \%) \\ 
\midrule
$f_s$=96000 Hz & $\Delta=0.01$ m & 40000 (100 \%) & 12362 (30.91 \%) & 31908 (79.77 \%) & 38358 (95.90 \%) & 39103 (97.76 \%) \\ 
 & $\Delta=0.05$ m& 1600 (100 \%) & 1512 (94.50 \%) & 1535 (95.94 \%) & 1548 (96.75 \%) & 1552 (97.00 \%) \\
 & $\Delta=0.1$ m& 400 (100 \%) & 374 (93.50 \%) & 380 (95.00 \%) & 380 (95.00 \%) & 380 (95.00 \%) \\
\bottomrule
\end{tabular}
\end{table*}

\begin{figure*}[tp]
\begin{minipage}[l]{1.0\columnwidth}
\centering
\includegraphics[width=\columnwidth]{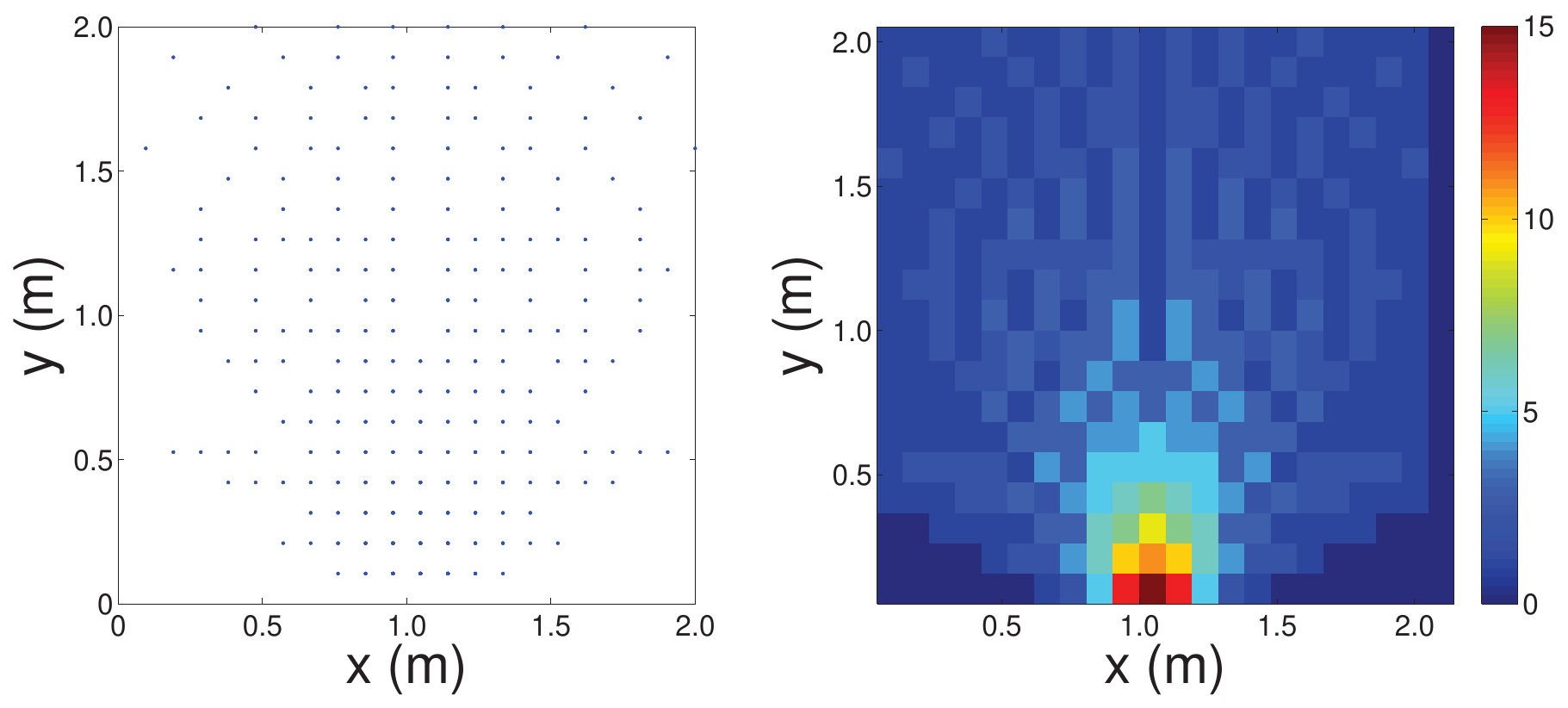}
\caption{The grid map $\Gamma_r$ and the sensitivity map $\delta(\mathbf{r}_g)$ for an ULA of 3 microphones, a space resolution $\Delta=0.1$ m and $f_s=16$ kHz.}\label{ula_16_01_3}
\end{minipage}
\hfill{}
\begin{minipage}[r]{1.0\columnwidth}
\centering
\includegraphics[width=\columnwidth]{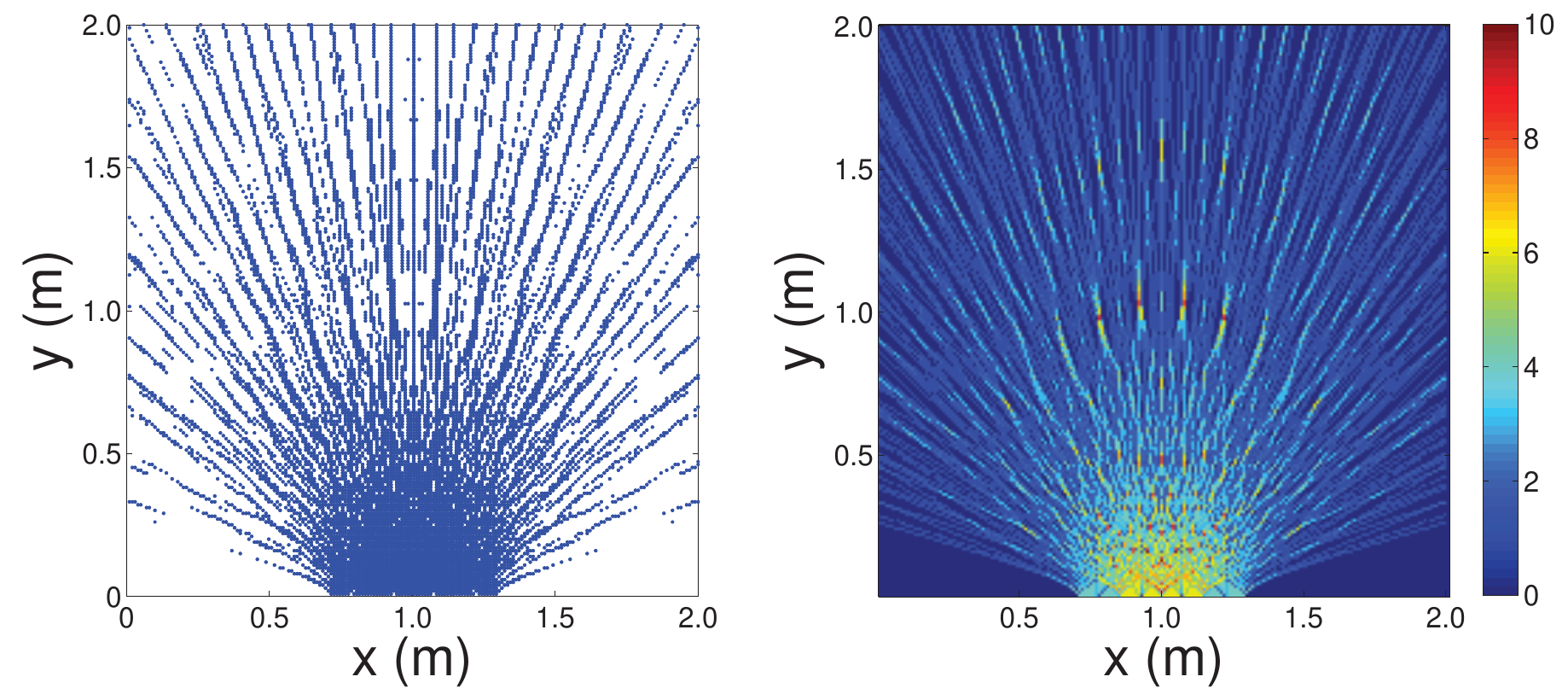}
\caption{The grid map $\Gamma_r$ and the sensitivity map $\delta(\mathbf{r}_g)$ for an ULA of 5 microphones, a space resolution $\Delta=0.01$ m and $f_s=16$ kHz.}
\label{ula_16_001_5}
\end{minipage}
\hspace{5mm}
\begin{minipage}[l]{1.0\columnwidth}
\centering
\includegraphics[width=\columnwidth]{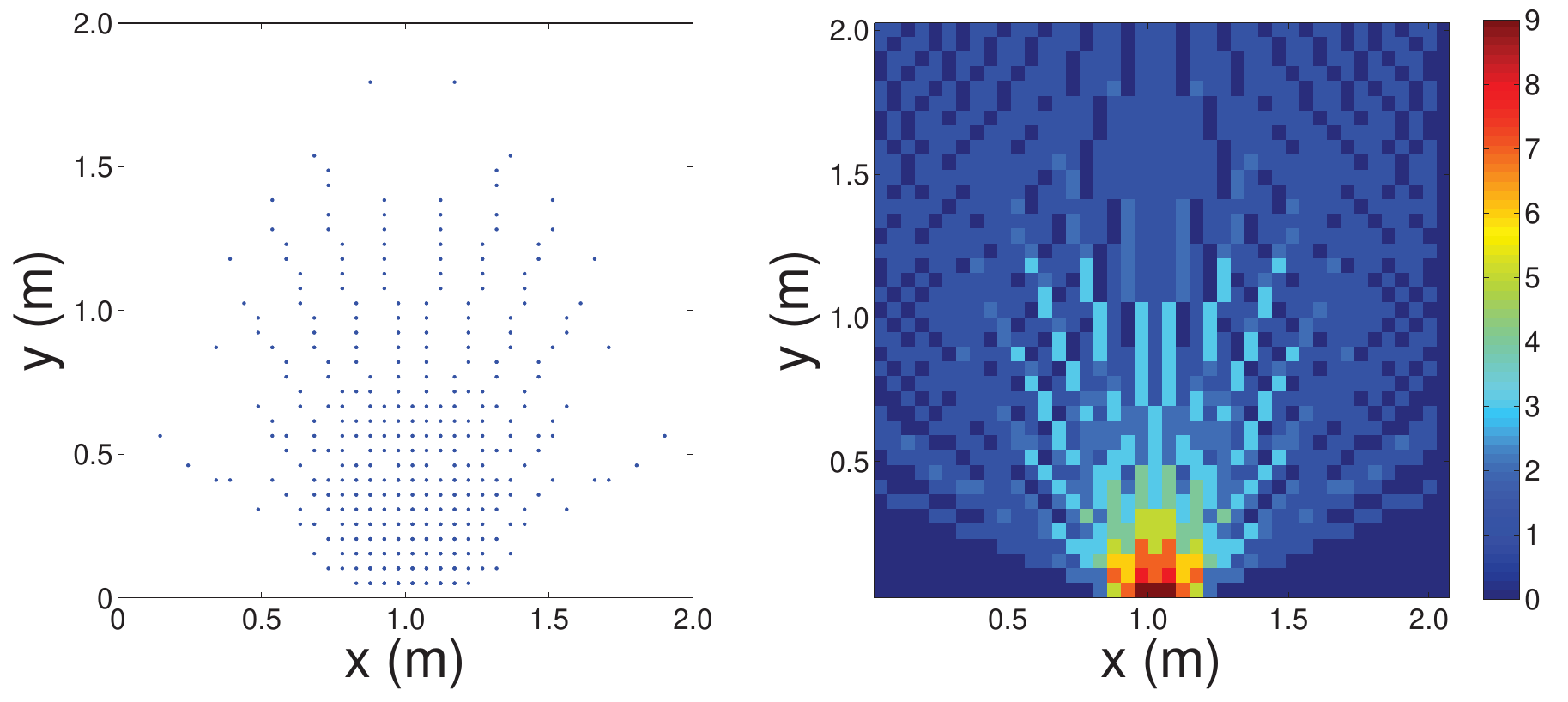}
\caption{The grid map $\Gamma_r$ and the sensitivity map $\delta(\mathbf{r}_g)$ for an ULA of 3 microphones, a space resolution $\Delta=0.05$ m and $f_s=16$ kHz.}
\label{ula_16_005_3}
\end{minipage}
\hfill{}
\begin{minipage}[r]{1.0\columnwidth}
\centering
\includegraphics[width=\columnwidth]{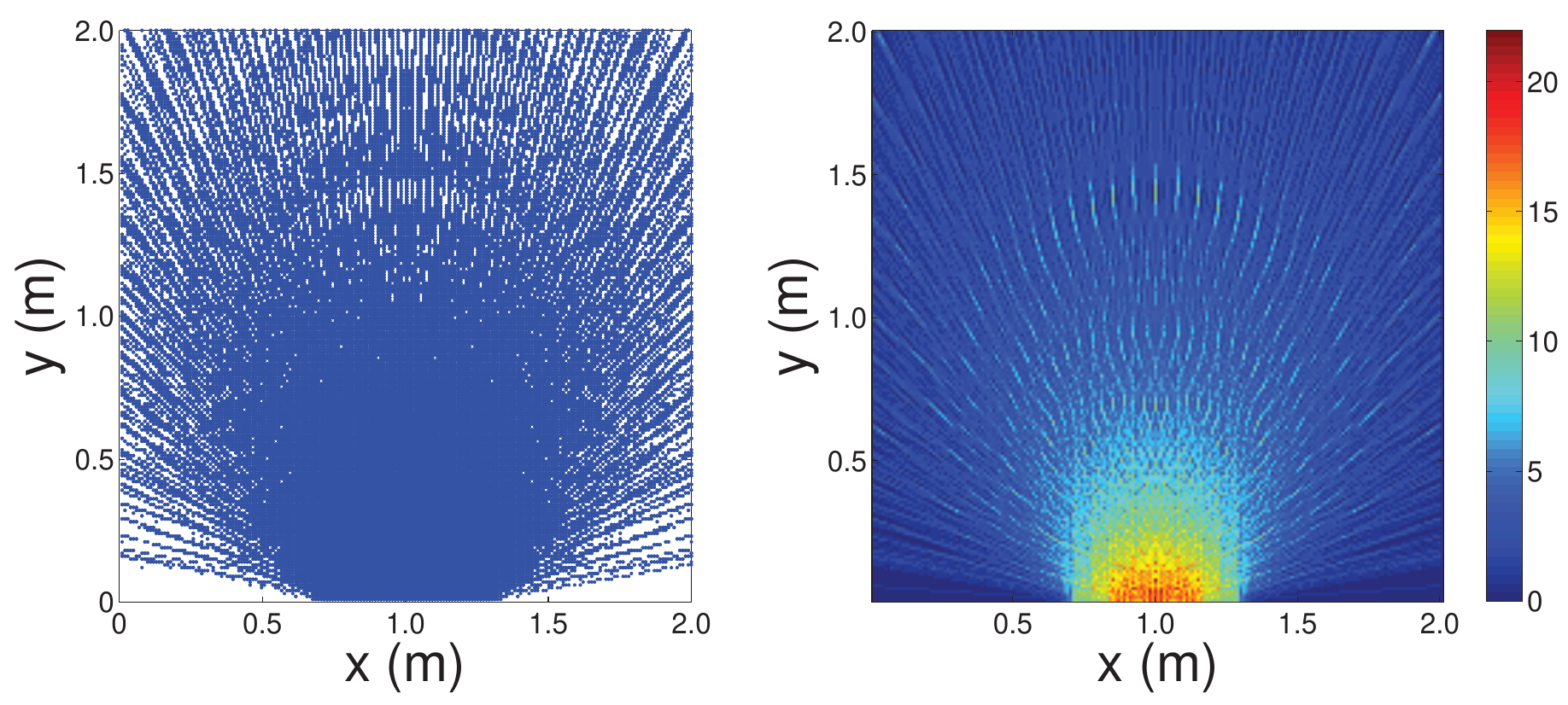}
\caption{The grid map $\Gamma_r$ and the sensitivity map $\delta(\mathbf{r}_g)$ for an ULA of 5 microphones, a space resolution $\Delta=0.01$ m and $f_s=44.1$ kHz.}
\label{ula_44_001_5}
\end{minipage}
\hspace{5mm}
\begin{minipage}[l]{1.0\columnwidth}
\centering
\includegraphics[width=\columnwidth]{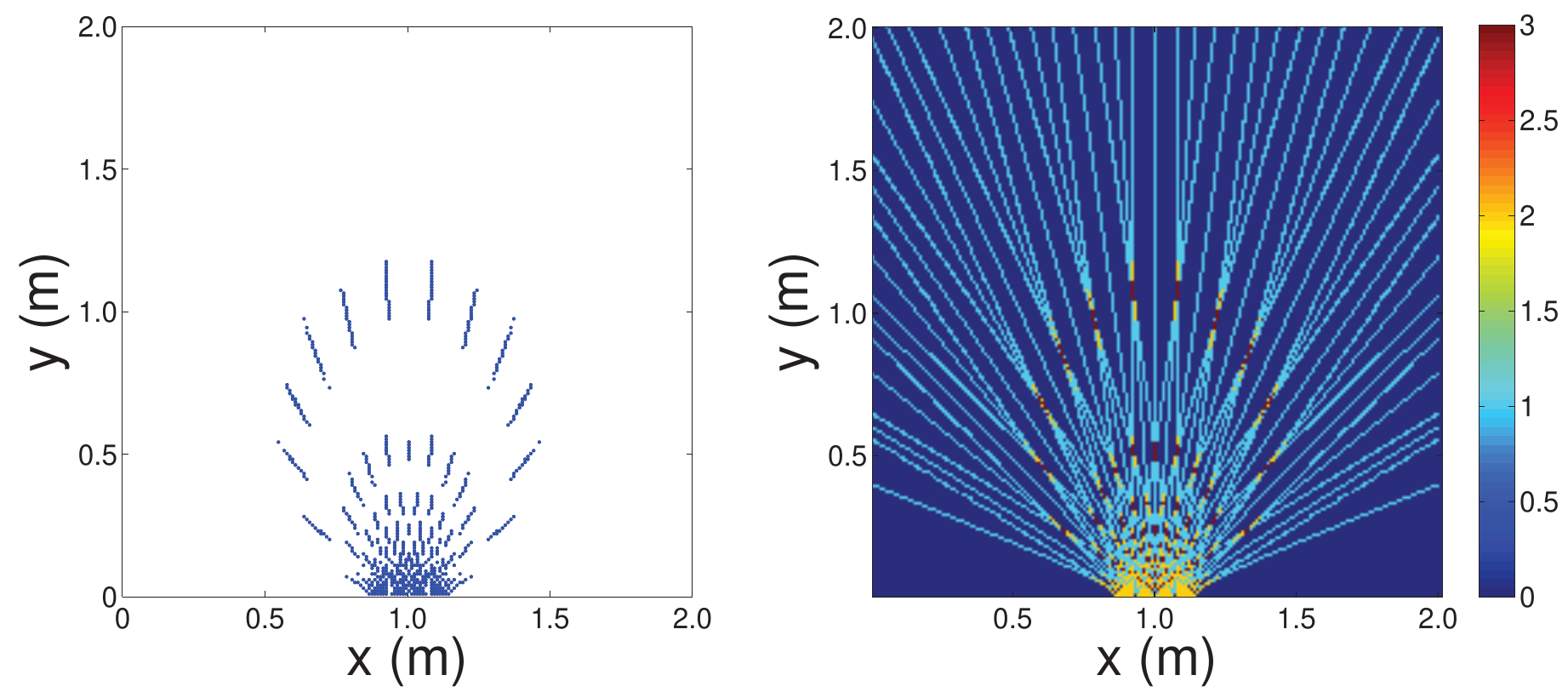}
\caption{The grid map $\Gamma_r$ and the sensitivity map $\delta(\mathbf{r}_g)$ for an ULA of 3 microphones, a space resolution $\Delta=0.01$ m and $f_s=16$ kHz.}
\label{ula_16_001_3}
\end{minipage}
\hfill{}
\begin{minipage}[r]{1.0\columnwidth}
\centering
\includegraphics[width=\columnwidth]{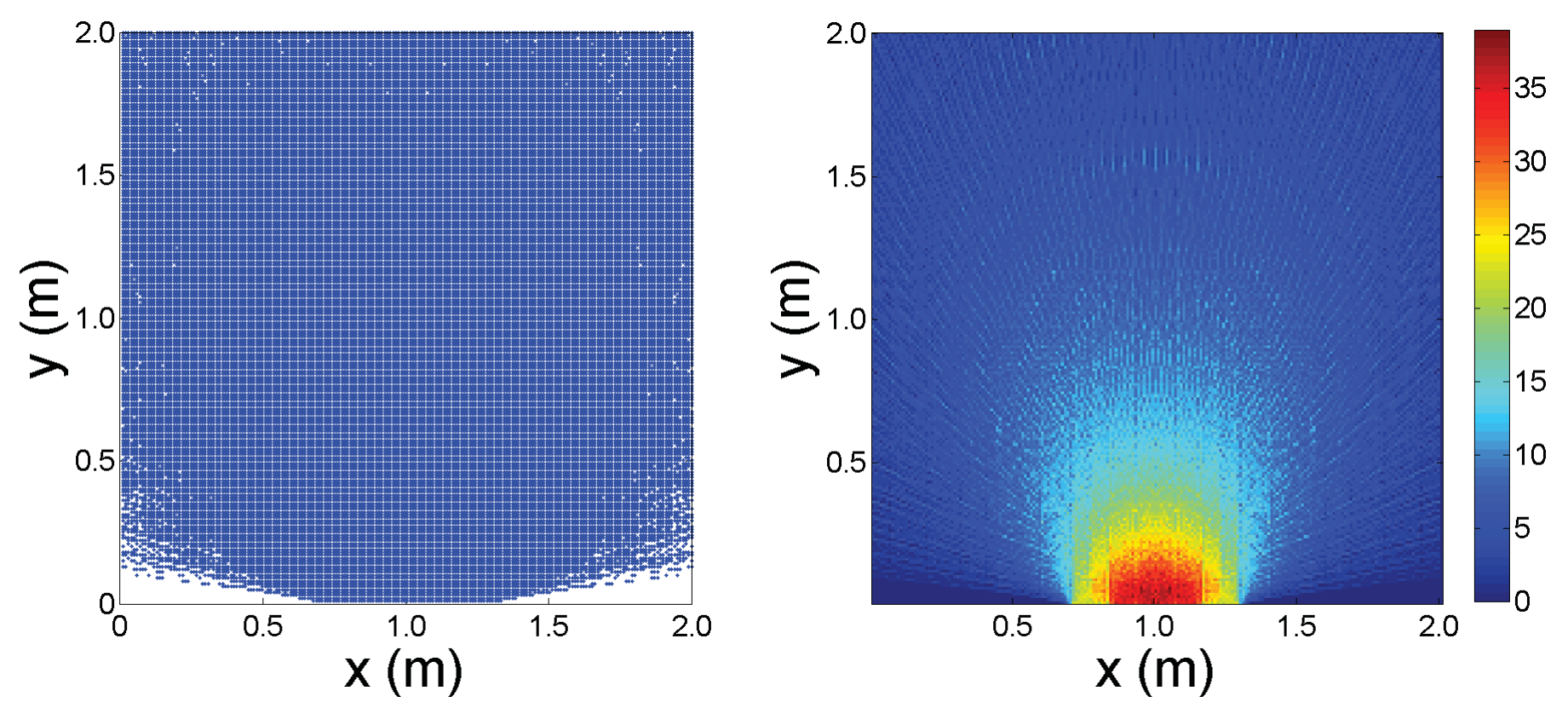}
\caption{The grid map $\Gamma_r$ and the sensitivity map $\delta(\mathbf{r}_g)$ for an ULA of 5 microphones, a space resolution $\Delta=0.01$ m and $f_s=96$ kHz.}
\label{ula_96_001_5}
\end{minipage}
\hspace{5mm}
\begin{minipage}[l]{1.0\columnwidth}
\centering
\includegraphics[width=1.0\columnwidth ]{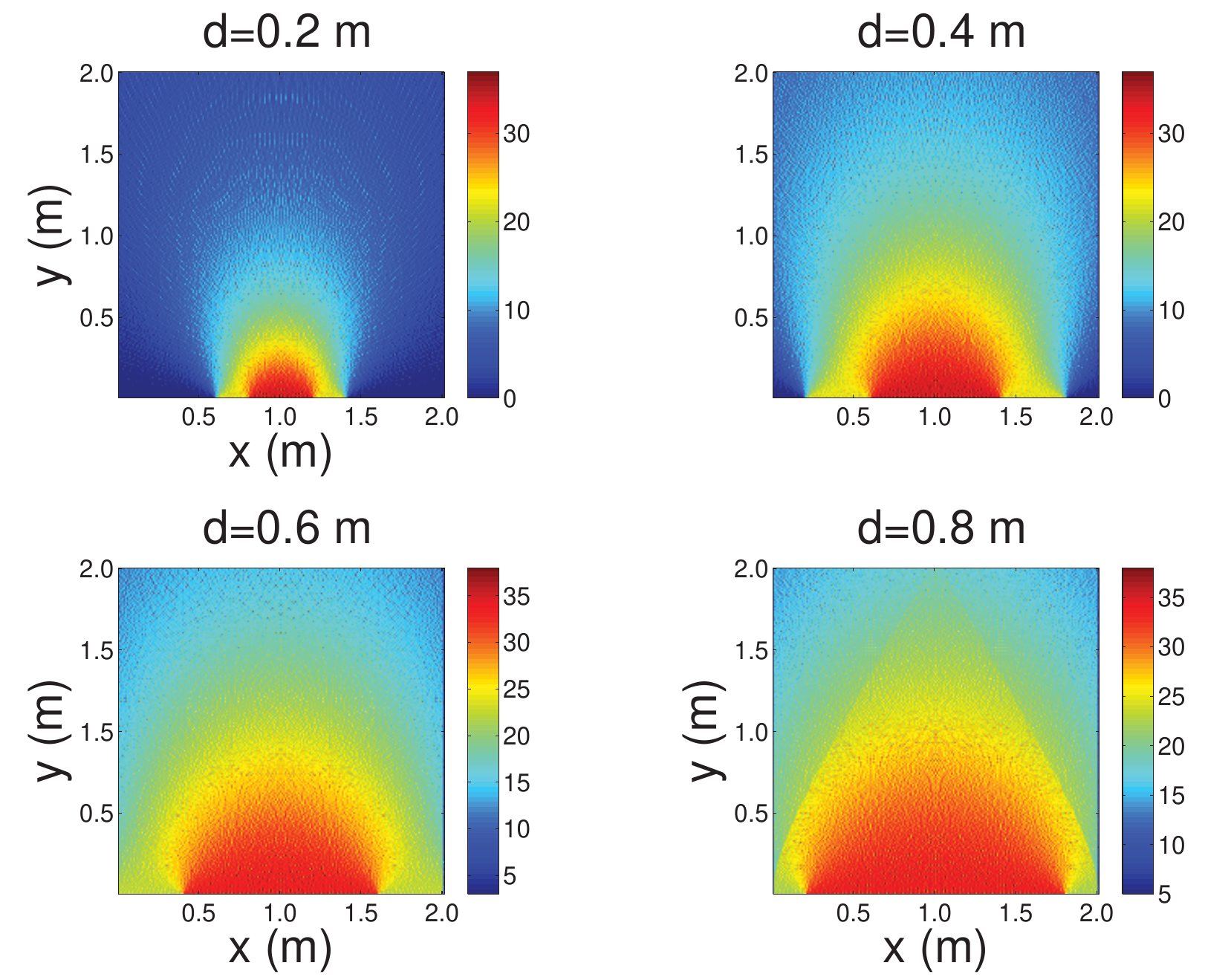}
\caption{The sensitivity map $\delta(\mathbf{r}_g)$ corresponding to four values of the inter-microphone distance $d$ for an ULA of 5 microphones, a space resolution $\Delta=0.01$ m and $f_s=96$ kHz.}
\label{smvd}
\end{minipage}
\hfill{}
\begin{minipage}[r]{1.0\columnwidth}
\centering
\includegraphics[width=1.0\columnwidth ]{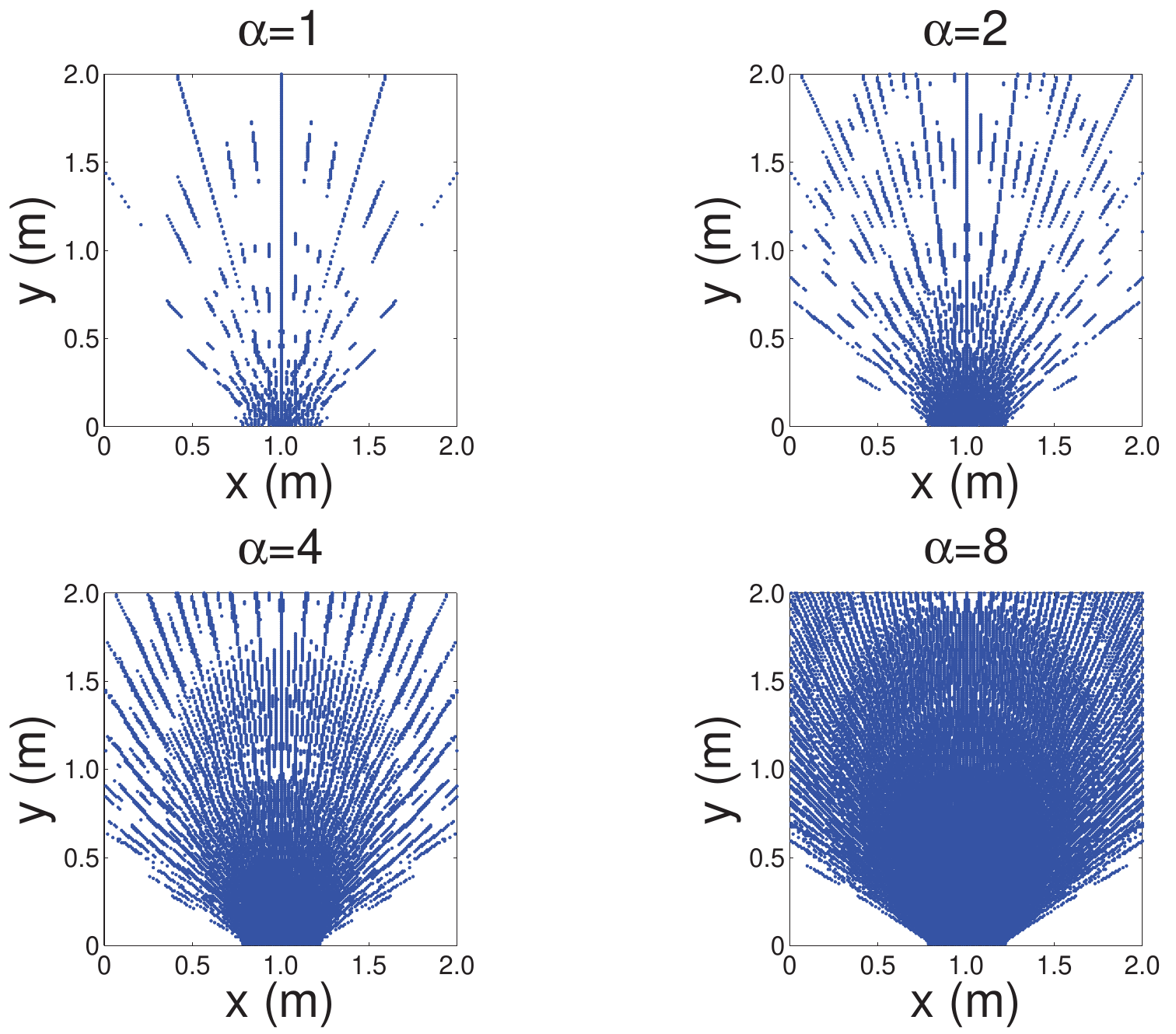}
\caption{The grid map $\Gamma_r$ with ($\alpha$=2,4,8) and without ($\alpha$=1) interpolation for an ULA of 4 microphones, a space resolution $\Delta=0.01$ m and $f_s=8$ kHz.}
\label{ula_8_001_upsampling}
\end{minipage}
\end{figure*}

Figures \ref{ula_16_01_3}, \ref{ula_16_001_5}, \ref{ula_16_005_3}, \ref{ula_44_001_5}, \ref{ula_16_001_3}, \ref{ula_96_001_5} depict the grid map $\Gamma_r$ and the sensitivity map $\delta(\mathbf{r}_g)$ calculated with the GSG algorithm for different system configurations. The center of the array is positioned at location (1,0) m. Note that the $\delta(\mathbf{r}_g)$ tables in the figures are reported before applying the constraint in equation (\ref{acon}). The colorbar on the right of the figures shows the number of the intersections of hyperbolas.

By observing the sensitivity maps, we can see how the GCC-PHAT functions are projected onto the search region, and how their values are accumulated. We note that the red colored regions are characterized by a high power response sensitivity since they accommodate a high number of hyperbola intersections. We can see in Figure \ref{ula_96_001_5} that the high sensitivity region accommodates a number of intersections contained in the range [25, 35], whereas the URG only accounts for $M(M-1)/2=10$ intersections at each point on the grid. Figure \ref{smvd} depicts the power response sensitivity analysis corresponding to different values of the array aperture, for an ULA of 5 microphones, a space resolution $\Delta=0.01$ m and $f_s=96$ kHz. We observe how the high sensitivity region (red-colored region) expands when the distance between microphone increases, due to the higher resolution of the GCC-PHAT functions that provide a larger number of hyperbolas for each sensor pair. 

The coherent spatial grid and the sensitivity map can be optimally constructed for a specific search region by properly configuring the geometry of the array, the number of microphones, and the sampling frequency. An alternative way to increase the TDOA resolution, and accordingly the number of hyperboloid of a sensor pair, is by interpolation. If $1/\alpha$ is an upsampling step, the possible TDOA values for the sensor pair $n$ will become $2\alpha T_n+1$. When interpolation is considered in the GSG, we have to calculate discrete hyperboloids also for non-integer TDOA values according to the parameter $\alpha$. An example of interpolation in the GSG is shown in Figure \ref{ula_8_001_upsampling}, in which we can observe the spatial grid corresponding to different values of $\alpha$, for an ULA of 4 microphones, a space resolution $\Delta=0.01$ m and $f_s=8$ kHz. Note that the effectiveness of interpolation for incrementing the spatial resolution is related to the signal-to-noise ratio (SNR) of the signal, and upsampling may lead to poor accuracy for low SNR \cite{Zhang2006}.

\begin{figure}[t]
\centering
\includegraphics[width=1.0\columnwidth ]{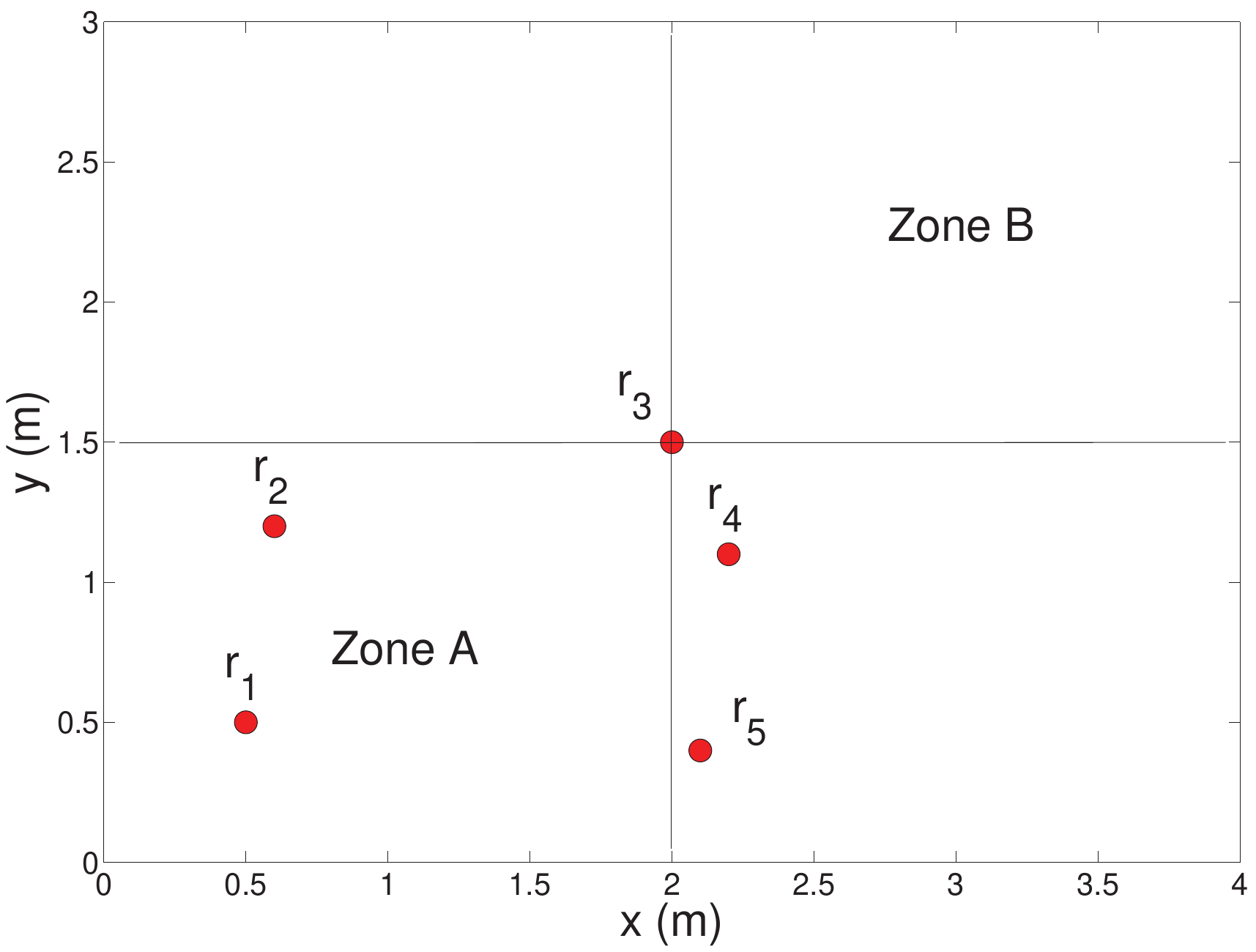}
\caption{The simulated room setup with the positions of the five microphones and the two zones A and B for evaluating the performance of SRP-PHAT with URG, URG-MSSS, URG-SSS, URG-VB and GSG algorithm. Two zones A and B were considered with high and low TDOA information taking into account the sensitivity map depicted in Figures \ref{amt2gsg001}, \ref{amt2gsg005}, and \ref{amt2gsg05}.}
\label{rst2new}
\end{figure}

In next sections, we will see the importance of the power response sensitivity analysis and how it is deeply related to the performance of sound source localization. 

\subsection{Localization Performance for Simulated Data}

In this section, the localization performance of the proposed GSG algorithm is assessed on a set of acoustic data simulated numerically. We also show that the sensitivity map obtained with the GSG algorithm is a useful tool to classify the areas in terms of high or poor localization performance. Besides that, we compare the performance of SRP-PHAT using URG \cite{DiBiase2001}, URG-SSS \cite{Cobos2011}, URG-MSSS \cite{Marti2013}, URG-VB \cite{Nunes2014} and GSG algorithm for different spatial resolution conditions: low $\Delta=0.5$ m, medium $\Delta=0.05$ m, and high $\Delta=0.01$ m.  

In the experiments with simulated acoustic data, a randomly distributed microphone network of 5 sensors was used. The image-source method (ISM) was used to simulate reverberant audio data in room acoustics \cite{Lehmann2008}. The ISM assumes that source and microphones are omnidirectional; it provides an approximation of the acoustic energy decay in room impulse responses generated using the image-source technique, and the sound sources are filtered through the impulse responses to produce reverberant signals. A localization task in two-dimensions, in a room of 4 m $\times$ 3 m $\times$ 3 m, was considered.
Therefore both microphones and the source were positioned at a distance from the floor of 1.7 m. The room setup is shown in Figure \ref{rst2new}. 

\begin{figure}[t]
\centering
\includegraphics[width=1.0\columnwidth ]{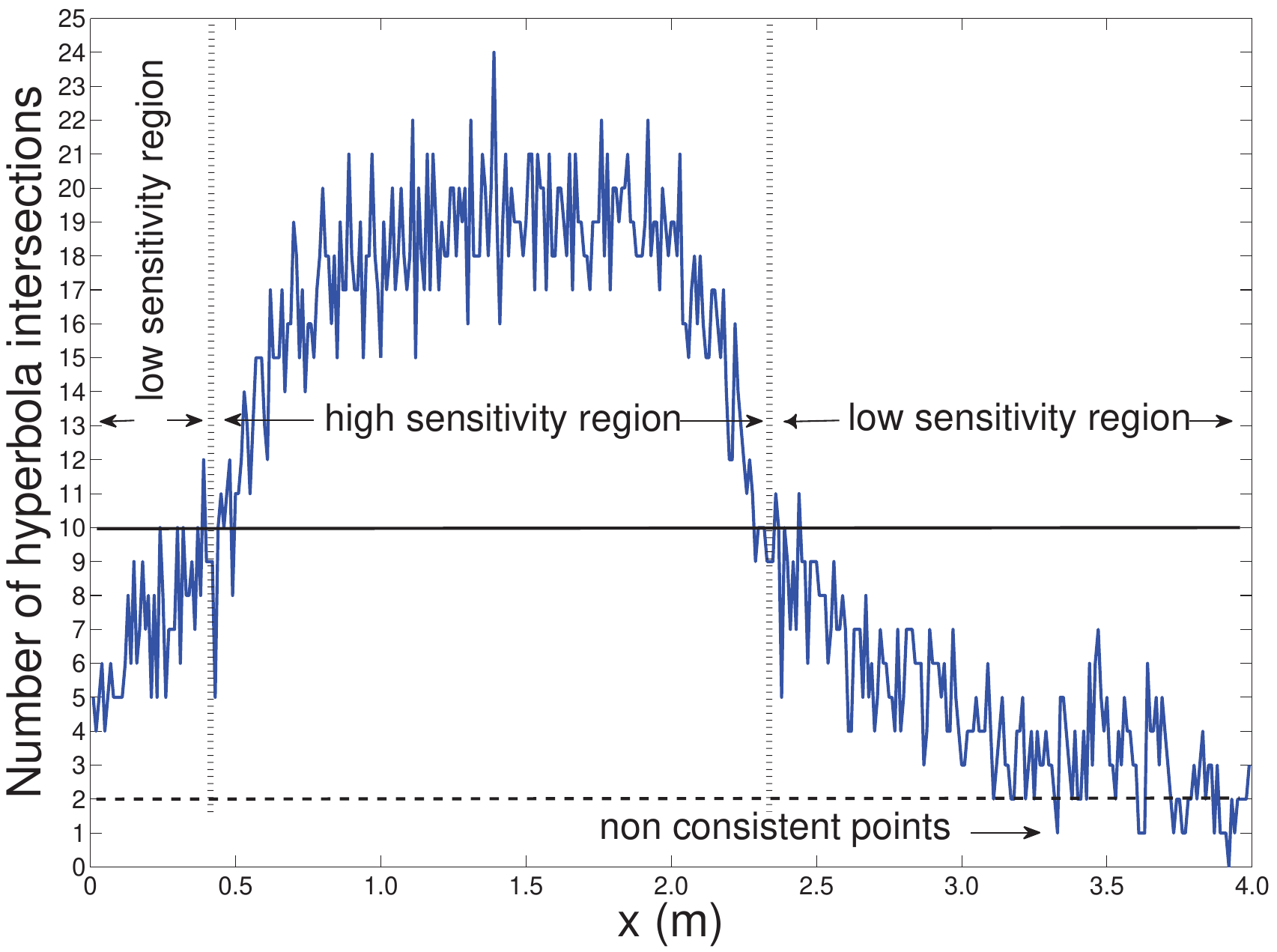}
\caption{The sensitivity response measure along $x$ axes for a $\Delta$ of 0.01 m and $y=1$ m. The horizontal solid line represents the number of hyperbola intersections assumed by the URG (10, if the number of sensors is 5 as in this case), and the horizontal dashed line represent
the minimum number of intersections for acoustical consistency (2, for 2D localization as in this case).}
\label{num_ype}
\end{figure}

\begin{figure*}[tp]
\begin{minipage}[l]{1.0\columnwidth}
\centering
\includegraphics[width=0.8\columnwidth]{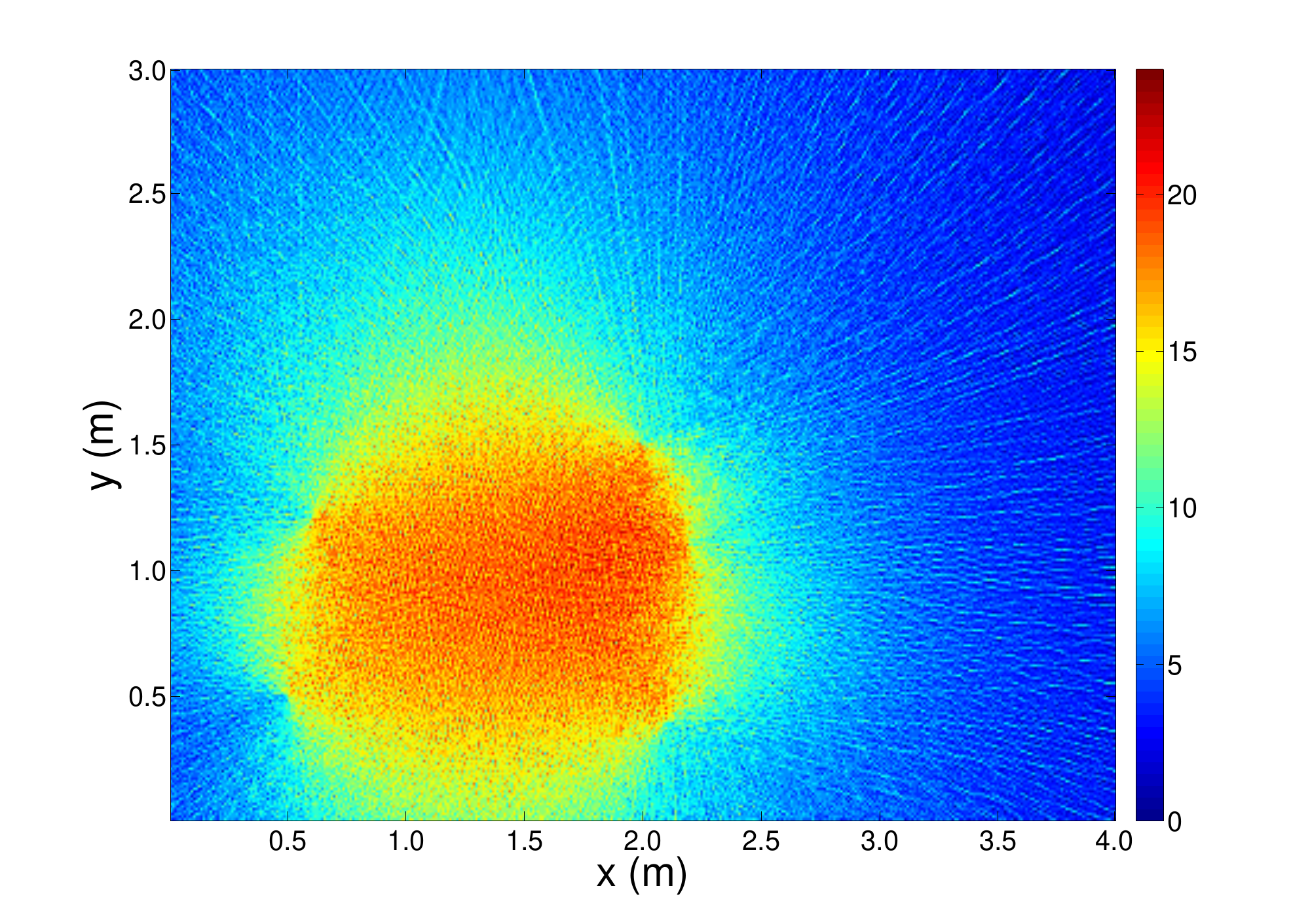}
\caption{The sensitivity map provided by the GSG table $\delta({\mathbf{r}_g})$ of the array in Figure \ref{rst2new} with  $\Delta=0.01$ m.}
\label{amt2gsg001}
\end{minipage}
\hfill{}
\begin{minipage}[r]{1.0\columnwidth}
\centering
\includegraphics[width=0.8\columnwidth]{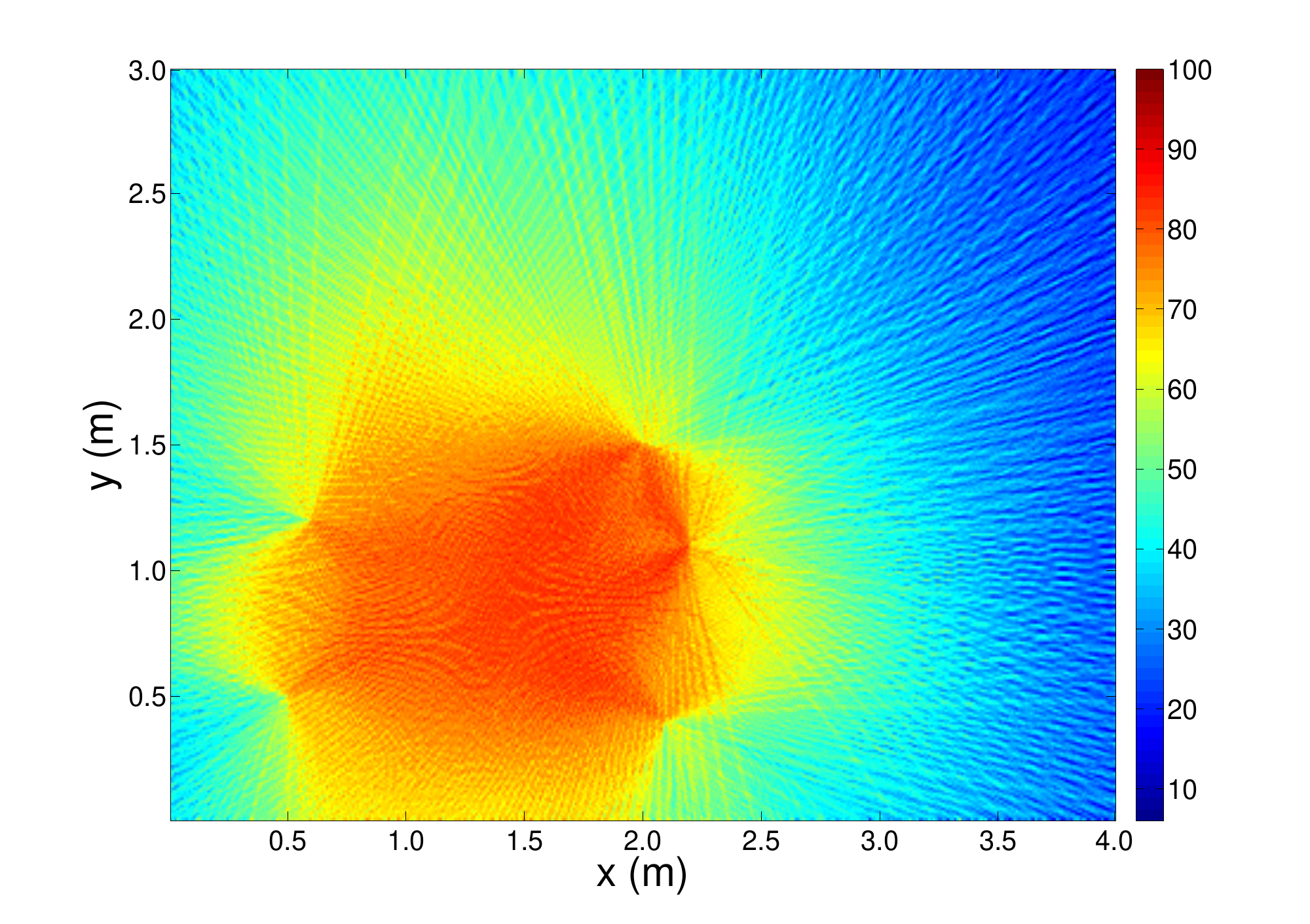}
\caption{The discriminability measure map \cite{Nunes2012} of the array in Figure \ref{rst2new} with  $\Delta=0.01$ m.}
\label{amt2dm001}
\end{minipage}
\hspace{5mm}
\begin{minipage}[l]{1.0\columnwidth}
\centering
\includegraphics[width=0.8\columnwidth]{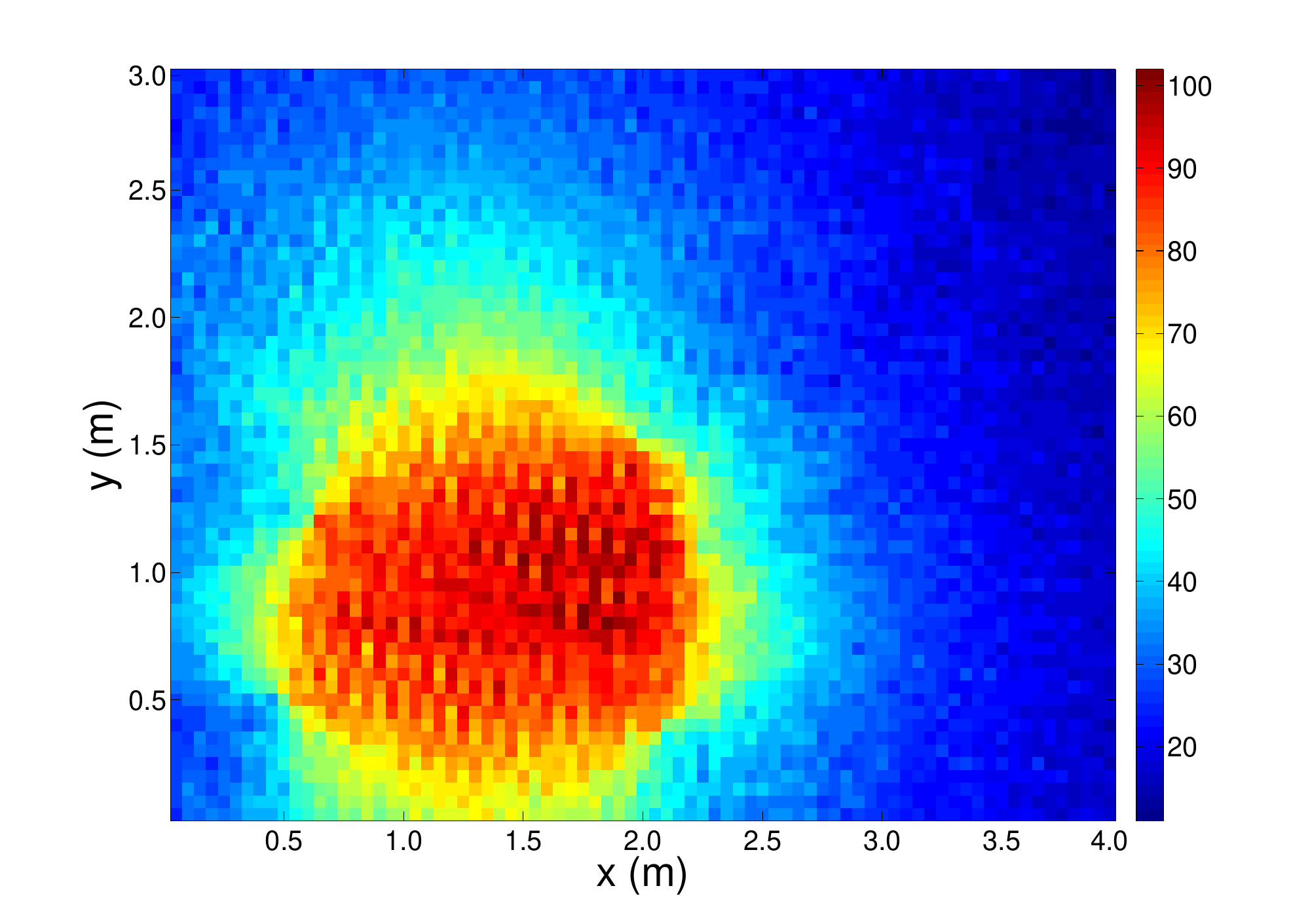}
\caption{The sensitivity map provided by GSG table $\delta({\mathbf{r}_g})$ of the array in Figure \ref{rst2new} with  $\Delta=0.05$ m.}
\label{amt2gsg005}
\end{minipage}
\hfill{}
\begin{minipage}[r]{1.0\columnwidth}
\centering
\includegraphics[width=0.8\columnwidth]{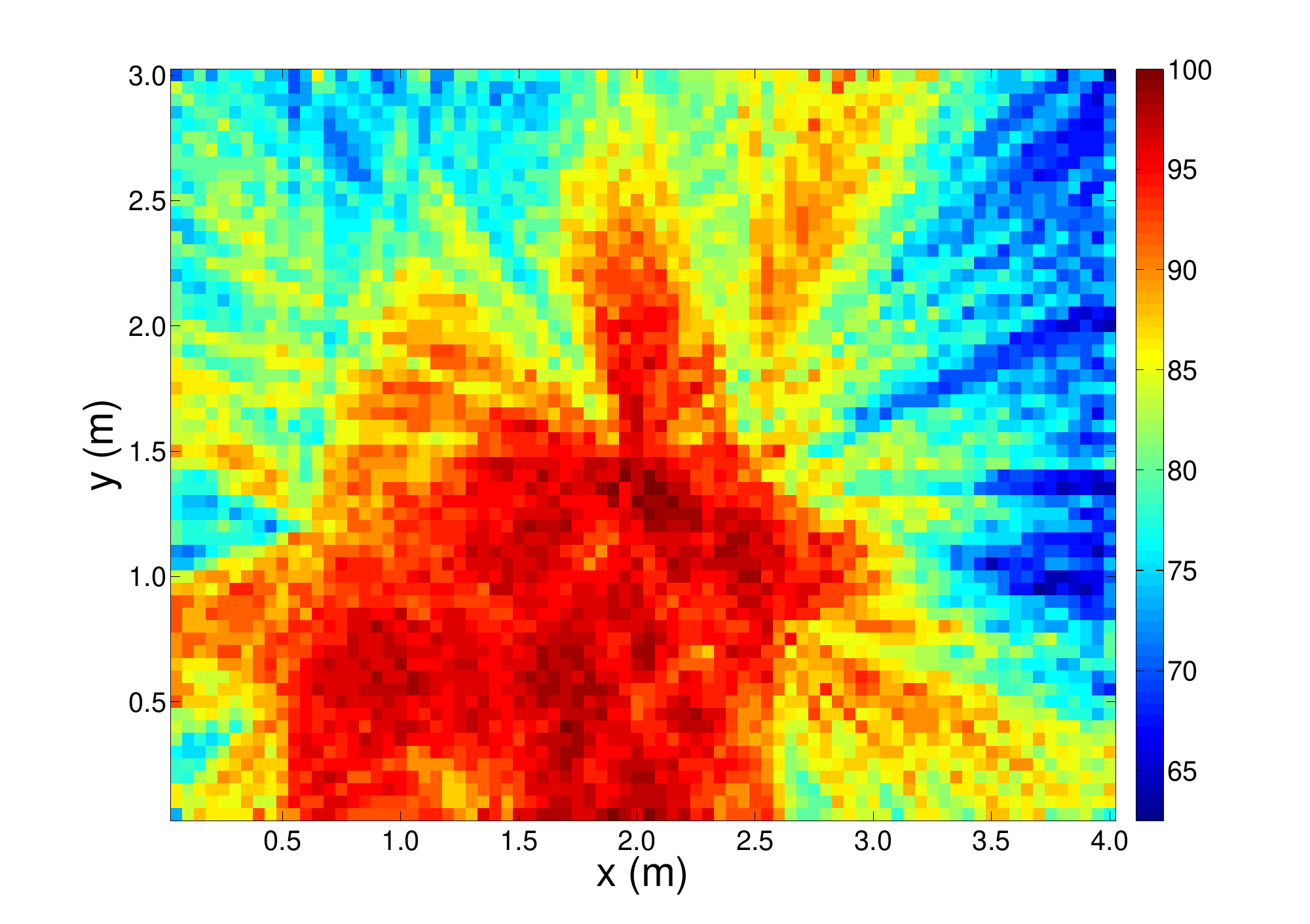}
\caption{The discriminability measure map \cite{Nunes2012} of the array in Figure \ref{rst2new} with  $\Delta=0.05$ m.}
\label{amt2dm005}
\end{minipage}
\hspace{5mm}
\begin{minipage}[l]{1.0\columnwidth}
\centering
\includegraphics[width=0.8\columnwidth]{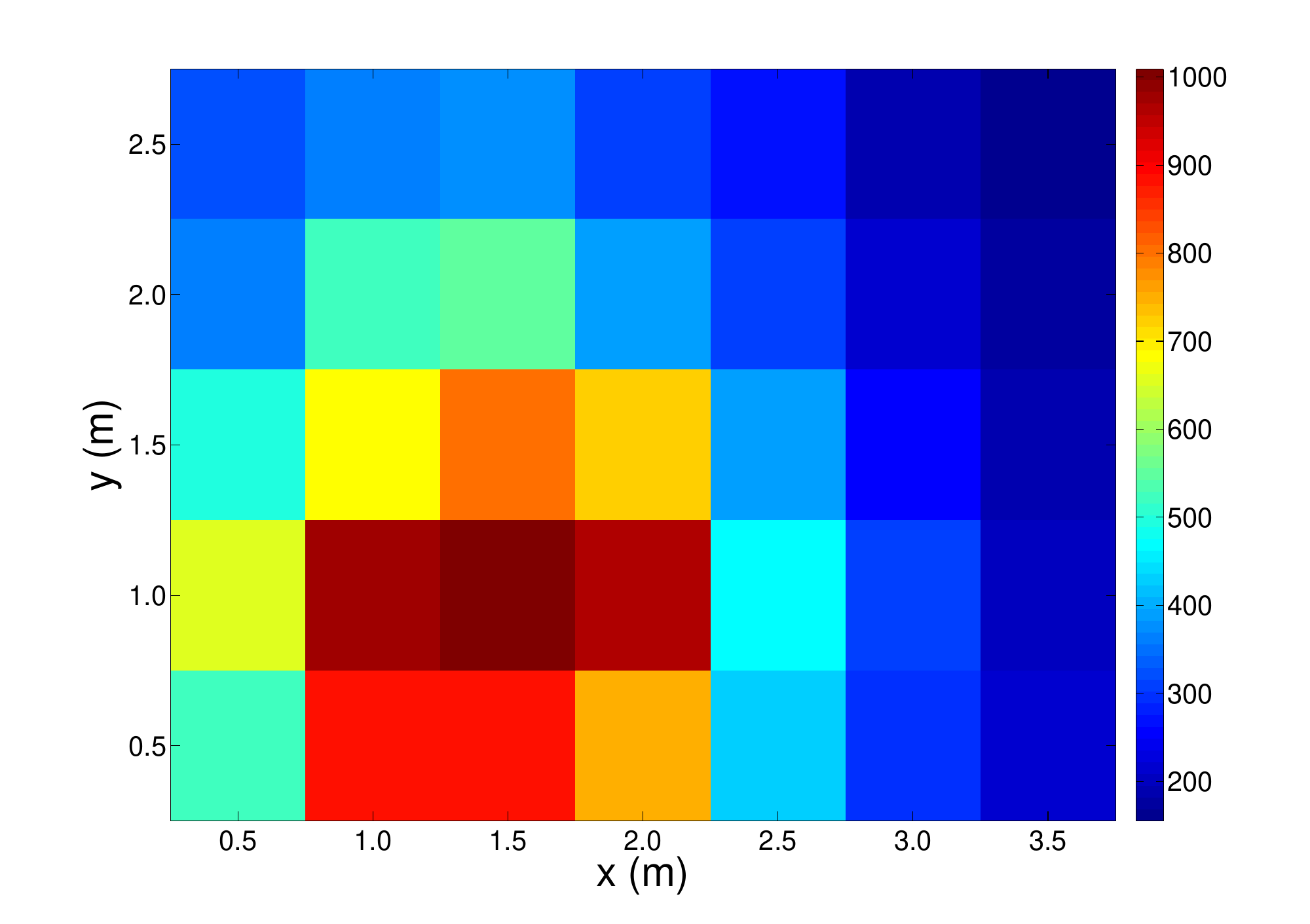}
\caption{The sensitivity map provided by GSG table $\delta({\mathbf{r}_g})$ of the array in Figure \ref{rst2new} with  $\Delta=0.5$ m.}
\label{amt2gsg05}
\end{minipage}
\hfill{}
\begin{minipage}[r]{1.0\columnwidth}
\centering
\includegraphics[width=0.8\columnwidth]{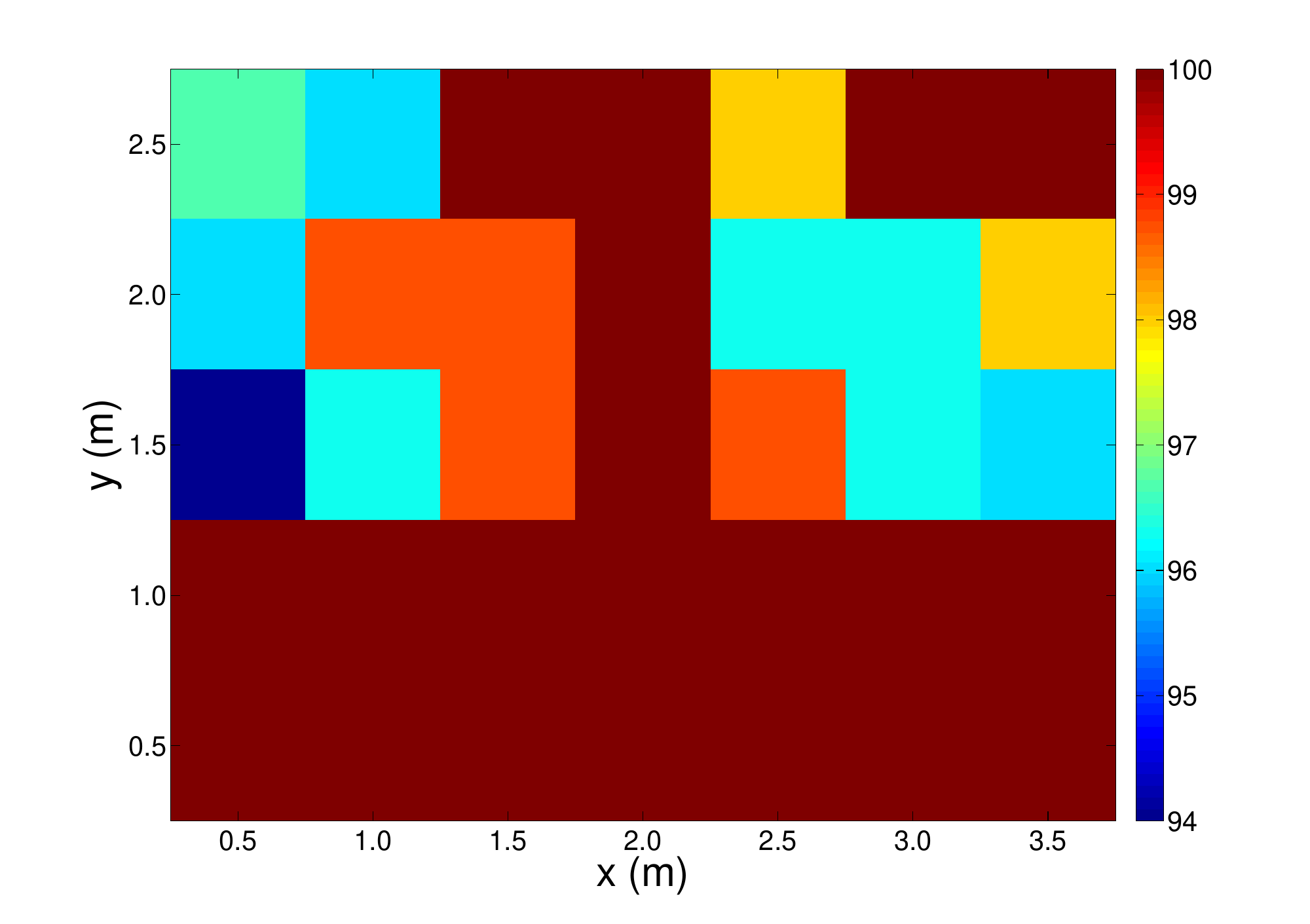}
\caption{The discriminability measure map \cite{Nunes2012} of the array in Figure \ref{rst2new} with  $\Delta=0.5$ m.}
\label{amt2dm05}
\end{minipage}
\end{figure*}

The $\delta$ table calculated with the GSG algorithm for a $\Delta$ of 0.01 m, of 0.05 m, and 0.5 m are depicted in Figures \ref{amt2gsg001}, \ref{amt2gsg005}, and \ref{amt2gsg05} respectively. We also report the discriminability measure map proposed in \cite{Nunes2012}. As we can observe in Figures \ref{amt2dm001}, \ref{amt2dm005}, and \ref{amt2dm05} the discriminability measure map is accurate for $\Delta=0.01$ m but it does not provide useful information for $\Delta=0.05$ m and $\Delta=0.5$ m, because of the TDOA information loss discussed so far. Figure \ref{num_ype} shows the sensitivity response measure in terms of hyperbola intersections along $x$ axes for a $\Delta$ of 0.01 m and $y=1$ m. The horizontal solid line represents the number of hyperbola intersections assumed by the URG. We note a greater number of intersections in the high sensitivity region with a range $x=[0.4;2.3]$.

The reverberant condition was set to 0.3 s and 0.9 s reverberation time (RT$_{60}$). A 25 s duration adult male speech was used as a source signal. The tests were conducted by setting a SNR of 10 dB, which was obtained by adding mutually independent white Gaussian noise to each channel. The sampling frequency was 44.1 kHz, the block size $L$ was 4096 samples. 

Two zones A and B were considered with high and low TDOA information, taking into account the sensitivity map depicted in Figures \ref{amt2gsg001}, \ref{amt2gsg005}, and \ref{amt2gsg05}. The performance of localization has been evaluated with several Monte Carlo simulations, using 100 run-trials for each condition test. The source was randomly positioned at each trail, at a minimum distance of 0.1 m from the walls and microphones. Performance is reported in terms of the percentage of accuracy rate (AR) estimated for those square errors that are less than a root mean square (RMS) error of 0.2 m, and by the RMS error for all the estimates. 

The localization performance is given in Table \ref{sim2}. First, we can observe that SRP-PHAT-GSG outperforms SRP-PHAT-URG in all test conditions for Zone A. Besides that, we note a rapid degradation of SRP-PHAT-URG performance when the spatial resolution decreases, while SRP-PHAT-GSG is more robust due to the improved TDOA information exploitation.  Then, note also that the number of grid points for GSG is the same of URG when $\Delta=0.1$ m and $\Delta=0.05$ m. However, in the case of $\Delta=0.01$ m the GSG grid points are about 3\% less than the URG grid points, slightly reducing the computational cost for the maximum value search. The average performance of the URG-SSS and of the URG-VB is comparable to that of the GSG. Specifically, GSG has a better AR and RMS in coarser grids ($\Delta=0.1$ m and $\Delta=0.05$ m), due to the use of all TDOA information that ensures a larger number of hyperbola intersections in the high sensitivity region.  URG-SSS and URG-VB provide instead better performance  when $\Delta=0.01$ m. In this case, the use of a fine grid reduces the accumulation of GSG. However, URG-SSS and URG-VB provide no clues to select the region with best localization accuracy, while GSG includes the sensitivity analysis, which gives important clues on how the whole TDOA information is distributed. In fact, in the low accuracy Zone B, all algorithms perform the localization with higher error if compared to Zone A. When reverberation time increases, the noisier condition degrades the GCC-PHAT performance and the poor TDOA information in that region makes the localization very difficult. In particular, GSG, URG-SSS, and URG-VB are affected by a consistent performance degradation due to the fact that in Zone B a low energy peak related to the acoustic source is subject to be masked by high energy noise peaks with high probability. This observation suggests that a zone selection procedure that gives information on which is the most promising searching area may help in increasing the localization performance of GSG, URG-SSS, and URG-VB in low level sensitivity zones. The URG-MSSS provides worse localization performance for Zone A if compared to that of GSG, URG-SSS, and URG-VB, due to the averaging of the GCC-PHAT for each volume of the search grid. 

\begin{table*}[!t]
\centering
\renewcommand{\arraystretch}{1.1}
\caption{RMS (\textrm{\normalfont m}) and AR (\%) (RMS$<$0.2 \textrm{\normalfont m}) of localization performance for SRP-PHAT with GSG, URG, URG-MSSS, URG-SSS, URG-VB in a simulated reverberant room using a speech signal and a SNR of 10 \textrm{\normalfont d}B.}
\label{sim2}
\centering
\begin{tabular}{@{}l|l|l|l|ccccc@{}}
\toprule
\multicolumn{4}{c|}{} & GSG & URG & URG-MSSS & URG-SSS & URG-VB\\
\midrule
RT$_{60}$=0.3 s & $\Delta=0.5$ m& Zone A & RMS (m) & 0.600	& 1.679	& 1.536 &	0.668 &	0.637\\
 &  & & AR (\%) & 38.76	& 6.32 &	12.97 &	35.55	& 35.30 \\
 & & Zone B & RMS (m)& 1.898	& 1.622 &	1.476 &	1.834	& 1.849 \\
 & & & AR (\%) & 1.14	& 3.92	& 6.19 & 2.39 & 1.66 \\
& $\Delta=0.05$ m& Zone A & RMS (m)& 0.292	& 1.224	& 1.564	& 0.310	& 0.315 \\
 &  & & AR (\%) & 87.79	& 48.00	& 58.67	& 87.25	& 86.57 \\
 & & Zone B & RMS (m) & 2.027	& 1.496	& 1.103	& 1.960	& 1.969 \\
 & & & AR (\%) & 6.91	& 30.29	& 38.01	& 13.29	& 12.75 \\
& $\Delta=0.01$ m& Zone A & RMS (m)& 0.257 & 0.665 & 1.262	& 0.243 & 0.229\\
 &  & & AR (\%) & 90.75  & 77.80 & 71.53 & 91.01 & 91.68\\
 & & Zone B & RMS (m)& 2.112 & 1.719 & 1.175 & 2.028 & 1.994\\
 & & & AR (\%) & 3.56 & 28.77 & 35.21	& 10.12 & 16.84\\
\midrule
RT$_{60}$=0.9 s & $\Delta=0.5$ m& Zone A & RMS (m) & 0.795	& 1.750	& 1.778	& 0.867	& 0.855 \\
 &  & & AR (\%) & 21.83	& 3.27	& 4.12	& 19.87	& 18.80 \\
 & & Zone B & RMS (m)& 2.063	& 1.771	& 1.775	& 2.045	& 2.057 \\
 &  & & AR (\%) & 0.27	& 2.06 & 	2.70	& 0.53	& 0.41 \\
& $\Delta=0.05$ m& Zone A & RMS (m)& 0.540	& 1.627	& 2.230	& 0.553	& 0.558\\
 & & & AR (\%) & 57.96	& 16.35	& 17.42	& 57.88	& 57.91 \\
 & & Zone B & RMS (m)& 2.177 &	1.917	& 1.569	& 2.168	& 2.170 \\
 & & & AR (\%) & 1.06	& 7.95	& 11.21	& 2.49	& 2.34 \\
& $\Delta=0.01$ m & Zone A & RMS (m)& 0.534 & 1.139	& 2.056	 & 0.547 & 	0.531\\
 &  & & AR (\%) & 61.93 & 40.86 & 31.06 & 62.90 & 65.32\\
 & & Zone B & RMS (m)& 2.138 & 2.078 & 1.592 & 	2.122 & 2.130\\
 & & & AR (\%) & 0.52 & 7.34 & 10.03 & 2.65 & 3.13\\
\bottomrule
\end{tabular}
\end{table*}

\subsection{Localization Performance for Real Data}

\begin{figure*}[tp]
\begin{minipage}[l]{1.0\columnwidth}
\centering
\includegraphics[width=\columnwidth]{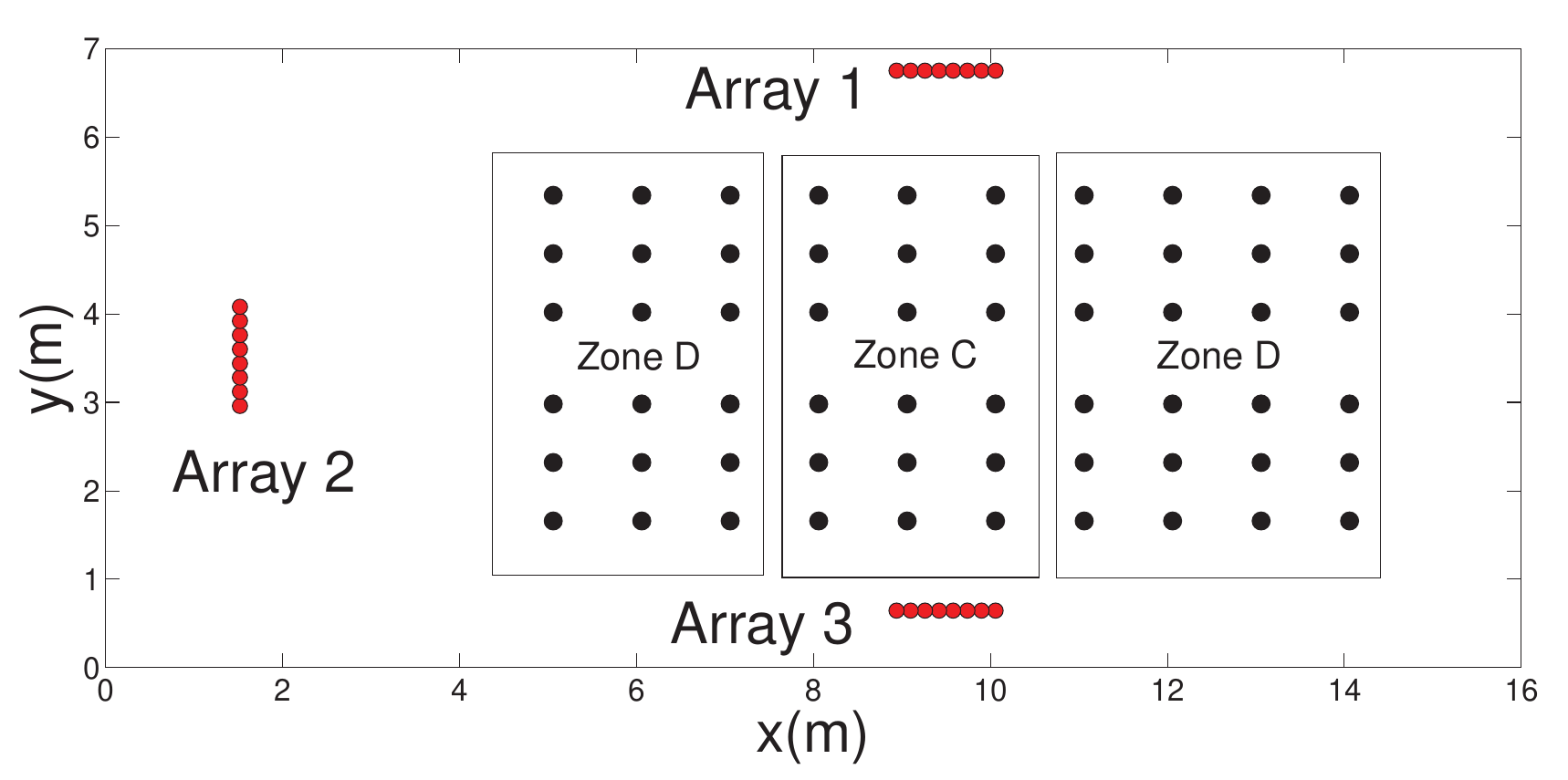}
\caption{The real-world room setup with the positions of the microphones and the speakers. Two zones C and D were considered with high and low TDOA information taking into account the sensitivity map depicted in Figures \ref{amrt}.}
\label{rsrt}
\end{minipage}
\hfill{}
\begin{minipage}[r]{1.0\columnwidth}
\centering
\includegraphics[width=\columnwidth]{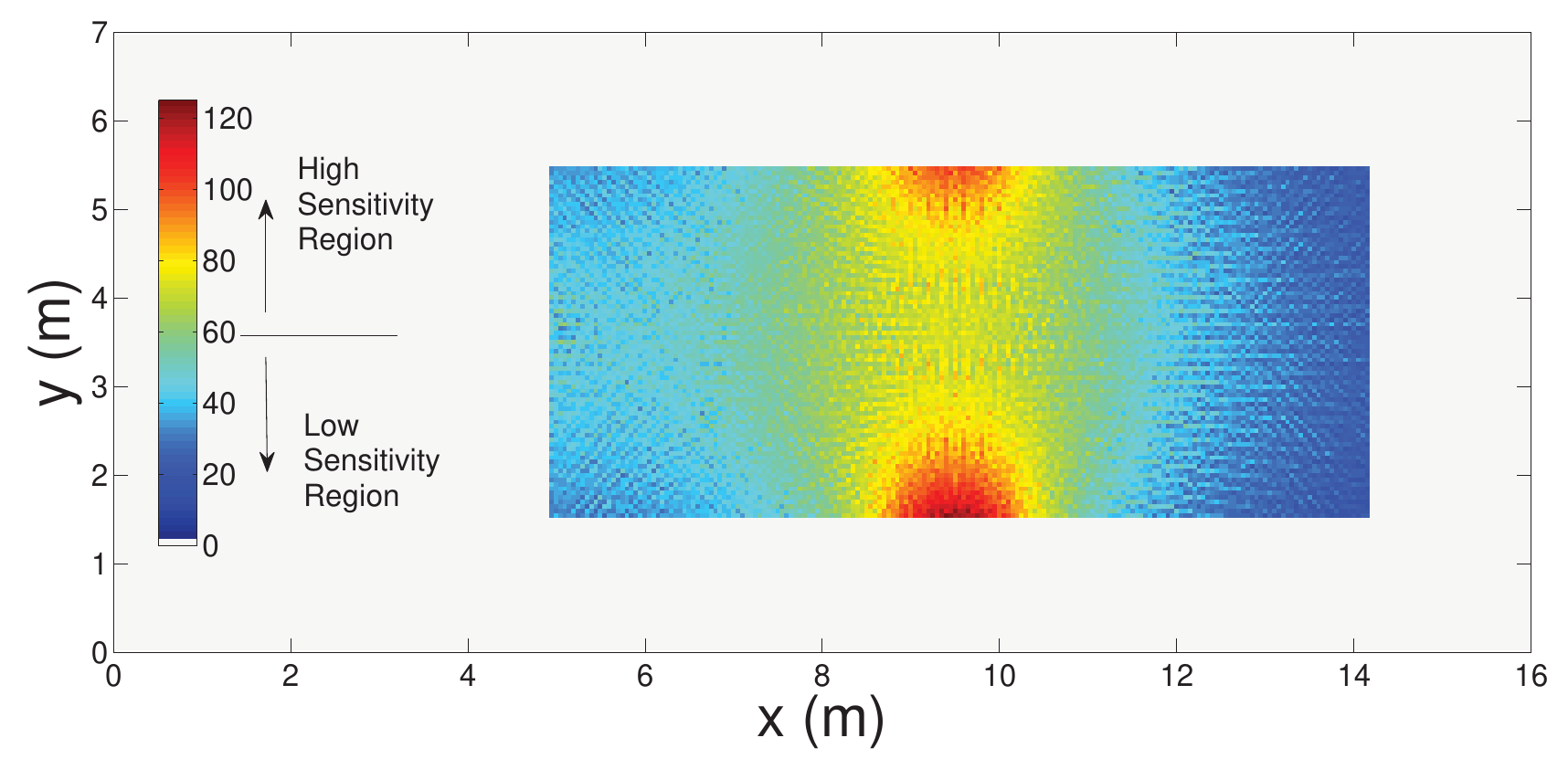}
\caption{The sensitivity map $\delta(\mathbf{r}_g)$ of the array in Figure \ref{rsrt} with $\Delta=0.05$ m and $f_s=48$ kHz.}
\label{amrt}
\end{minipage}
\end{figure*}

We report extensive tests computed in a real-world setup. An acoustic sensor network of 24
microphones has been installed in a conference room equipped with various multimedia facilities. 
The net of microphones is composed of 3 arrays, each one composed by 8 microphones arranged in a ULA with a distance between sensors of 0.16 m. The arrays are positioned with a distance from the floor of 1.7 m. The room setup is showed in Figure \ref{rsrt}, which reports also the source position (black circles) that has been used during recordings. The room dimensions in the x, y, z coordinates was 16 m $\times$ 7 m $\times$ 3 m, and its measured reverberation time was approximately 0.9 s of RT$_{60}$. The high reverberation time is due to the presence of glass window panes on the two sidewalls of the room. We have considered a position search area of dimensions 9.2 m $\times$ 3.88 m, and the $\delta$ table was calculated with the GSG algorithm for an imposed spatial resolution $\Delta$ of 0.05 m. The resulting sensitivity map $\delta({\mathbf{r}_g})$ is depicted in Figure \ref{amrt}. The grid points calculated with the GSG algorithm cover all the localization area, i.e, they are equal to URG in this specific case. All microphone pairs of each array has been used so that $N=84$. We have defined two zones (see Figure \ref{rsrt}) for evaluating the localization performance taking into account the sensitivity map depicted in Figure \ref{amrt}: a high sensitivity region (Zone C) and a low sensitivity region (Zone D). 

A speech database was recorded in the conference room to design and tune the acoustic localization front-end of the system. 
Collected data consisted of a sequence of short sentences uttered by two male and one female speakers, standing up at different positions in the room showed in Figure \ref{rsrt} with black circles. 
The recordings were organized in ten sessions, in which one speaker for each session changed four to eight locations, each time repeating his new position in the room. The total database consists of about 30 minutes of audio. The 24-channel audio was acquired at 48 kHz. The SRP-PHAT was computed with a block size $L$ of 4096 samples,  a overlap step of $L/4$. The parameters are evaluated in terms of AR percentage estimates for RMS$<$0.2 m, and overall RMS error. 

Table \ref{rt} shows the obtained results for the two zones. As we can see, the localization performance of all algorithms is more robust in terms of RMS error and AR in the high sensitivity region (Zone C), and we can observe the decrease of performance of all algorithms when the source was positioned in the low sensitivity region (Zone D). Note that the distinction between high-sensitivity and low-sensitivity areas in the search space is less marked than it was in the simulated experiments. Actually, the most of Zone C turns out to be characterized by a midrange valued sensitivity map, as we can see in Figure \ref{amrt}, and the areas with greater sensitivity are positioned near the arrays 1 and 3 (red zones). Thus, the performance gap between URG, URG-MSSS and GSG, URG-SSS, URG-VB is also less marked in comparison to the simulated experiments. Specifically, GSG has the best AR in the high sensitivity region, while URG-SSS and URG-VB has a slightly lower overall RMS.

\begin{table*}[!t]
\centering
\renewcommand{\arraystretch}{1.1}
\caption{RMS (\textrm{\normalfont m}) and AR (\%) (RMS$<$0.2 \textrm{\normalfont m}) of localization performance for SRP-PHAT with GSG, URG, URG-MSSS, URG-SSS, and URG-VB  in a real room with a RT$_{60}$ of 0.9 s.}
\label{rt}
\centering
\begin{tabular}{@{}l|l|ccccc@{}}
\toprule
\multicolumn{2}{c|}{} & GSG & URG & URG-MSSS & URG-SSS & URG-VB\\
\midrule
Zone C & RMS (m) & 1.267	& 1.737	& 1.986 &	1.134 &	1.161\\
 & AR (\%) & 32.42	& 22.34 &	22.39 &	27.53	& 26.41 \\
Zone D & RMS (m)& 3.428	& 2.799 &	3.011 &	2.789	& 2.699 \\
& AR (\%) & 7.65	& 9.82	& 10.60 & 10.06 & 11.40 \\
\bottomrule
\end{tabular}
\end{table*}

\section{Conclusions}

The paper proposes an algorithm for acoustic spatial grid design of the SRP-PHAT method. It is based on the geometry of discrete sampling of TDOA functions and the spatial resolution. The advantages of the GSG algorithm for the localization problem of an acoustic source in a reverberant environment are the following:

\begin{itemize}
\item It permits the calculation of a sensitivity map, which is a useful tool for identifying the best accuracy zone of a sensor array;
\item It allows the design of a spatial grid which is coherent with the acoustic information provided by the sensors array;
\item It links all sampling TDOA information from the GCC-PHAT functions into the space resulting in an improved localization in the high sensitivity region;
\item SRP-PHAT-GSG performance does not degrade when used with a low spatial resolution grid,  due to its spatial resolution scalability properties;
\item It permits the reduction of computational cost in those cases in which using the proposed spatial grid is appropriate for the given application or when restricting the search to an high accuracy area for localization;
\item It is a useful tool for the reconfiguration of the system, if the setup is not adequate to a specific target.
\end{itemize}

Experiments were conducted to show the coherent grid design and to analyze the power response sensitivity in case of small-size arrays at changing of system parameters: microphone number, sampling frequency, spatial resolution, and microphone distance. Next, by simulations and real-world experimental results, we have shown the importance of the steered response sensitivity analysis in the localization performance. We have demonstrated that high localization accuracy is achieved in the areas of
high sensitivity, while in the low sensitivity region the performance is degraded. Hence, GSG can be used to properly configure the array in order to let the higher
sensitivity zones maximally overlap with the target location area.

\balance

\bibliographystyle{IEEEtran}
\bibliography{gsg.v3}

\begin{thebibliography}{10}
\providecommand{\url}[1]{#1}
\csname url@samestyle\endcsname
\providecommand{\newblock}{\relax}
\providecommand{\bibinfo}[2]{#2}
\providecommand{\BIBentrySTDinterwordspacing}{\spaceskip=0pt\relax}
\providecommand{\BIBentryALTinterwordstretchfactor}{4}
\providecommand{\BIBentryALTinterwordspacing}{\spaceskip=\fontdimen2\font plus
\BIBentryALTinterwordstretchfactor\fontdimen3\font minus
  \fontdimen4\font\relax}
\providecommand{\BIBforeignlanguage}[2]{{%
\expandafter\ifx\csname l@#1\endcsname\relax
\typeout{** WARNING: IEEEtran.bst: No hyphenation pattern has been}%
\typeout{** loaded for the language `#1'. Using the pattern for}%
\typeout{** the default language instead.}%
\else
\language=\csname l@#1\endcsname
\fi
#2}}
\providecommand{\BIBdecl}{\relax}
\BIBdecl

\bibitem{omologo1998}
M.~Omologo, P.~Svaizer, and R.~{De Mori}, \emph{Spoken Dialogue with
  Computers}.\hskip 1em plus 0.5em minus 0.4em\relax Academic Press, 1998, ch.
  Acoustic Transduction.

\bibitem{DiBiase2001}
J.~H. DiBiase, H.~F. Silverman, and M.~S. Brandstein, \emph{Microphone Arrays:
  Signal Processing Techniques and Applications}.\hskip 1em plus 0.5em minus
  0.4em\relax Springer, 2001, ch. Robust localization in reverberant rooms.

\bibitem{Aarabi2003}
P.~Aarabi, ``The fusion of distributed microphone arrays for sound
  localization,'' \emph{EURASIP Journal on Applied Signal Processing}, vol.
  2003, no.~4, pp. 338--347, 2003.

\bibitem{Ward2003}
D.~B. Ward, E.~A. Lehmann, and R.~C. Williamson, ``Particle filtering
  algorithms for tracking an acoustic source in a reverberant environment,''
  \emph{IEEE Transactions on Speech and Audio Processing}, vol.~11, no.~6, pp.
  826--836, 2003.

\bibitem{Pertila2008}
P.~Pertil\"{a}, T.~Korhonen, and A.~Visa, ``Measurement combination for
  acoustic source localization in a room environment,'' \emph{EURASIP Journal
  on Audio, Speech, and Music Processing}, vol. 2008, pp. 1--14, 2008.

\bibitem{Velasco2012}
J.~Velasco, D.~Pizarro, and J.~Macias-Guarasa, ``Source localization with
  acoustic sensor arrays using generative model based fitting with sparse
  constraints,'' \emph{Sensors}, vol.~12, no.~10, pp. 13\,781--13\,812, 2012.

\bibitem{Knapp1976}
C.~Knapp and G.~Carter, ``The generalized correlation method for estimation of
  time delay,'' \emph{IEEE Transactions on Acoustics, Speech and Signal
  Processing}, vol.~24, no.~4, pp. 320--327, 1976.

\bibitem{Brutti2010}
A.~Brutti, M.~Omologo, and P.~Svaizer, ``Multiple source localization based on
  acoustic map de-emphasis,'' \emph{EURASIP Journal on Audio, Speech, and Music
  Processing}, vol. 2010, pp. 1--17, 2010.

\bibitem{Salvati2014}
D.~Salvati, C.~Drioli, and G.~L. Foresti, ``Incoherent frequency fusion for
  broadband steered response power algorithms in noisy environments,''
  \emph{IEEE Signal Processing Letters}, vol.~21, no.~5, pp. 581--585, 2014.

\bibitem{Berger1991}
M.~F. Berger and H.~F. Silverman, ``Microphone array optimization by stochastic
  region contraction,'' \emph{IEEE Transactions on Signal Processing}, vol.~39,
  no.~11, pp. 2377--2386, 1991.

\bibitem{Zotkin2004}
D.~N. Zotkin and R.~Duraiswami, ``Accelerated speech source localization via a
  hierarchical search of steered response power,'' \emph{IEEE Transactions on
  Speech and Audio Processing}, vol.~12, no.~5, pp. 499--508, 2004.

\bibitem{Dmochowski2007}
J.~P. Dmochowski, J.~Benesty, and S.~Affes, ``A generalized steered response
  power method for computationally viable source localization,'' \emph{IEEE
  Transactions on Audio, Speech and Language Processing}, vol.~15, no.~6, pp.
  2510--2526, 2007.

\bibitem{Nunes2014}
L.~O. Nunes, W.~A. Martins, M.~V.~S. Lima, L.~W.~P. Biscainho, M.~V.~M. Costa,
  F.~M. Gonçalves, A.~Said, and B.~Lee, ``A steered-response power algorithm
  employing hierarchical search for acoustic source localization using
  microphone arrays,'' \emph{IEEE Transactions on Signal Processing}, vol.~62,
  no.~19, pp. 5171--5183, 2014.

\bibitem{Cobos2011}
M.~Cobos, A.~Marti, and J.~J. Lopez, ``A modified {SRP-PHAT} functional for
  robust real-time sound source localization with scalable spatial sampling,''
  \emph{IEEE Signal Processing Letters}, vol.~18, no.~1, pp. 71--74, 2011.

\bibitem{Marti2013}
A.~Marti, M.~Cobos, J.~J. Lopez, and J.~Escolano, ``A steered response power
  iterative method for high-accuracy acoustic source localization,''
  \emph{Journal of the Acoustical Society of America}, vol. 134, no.~4, pp.
  2627--2630, 2013.

\bibitem{Lima2015}
M.~V.~S. Lima, W.~A. Martins, L.~O. Nunes, L.~W.~P. Biscainho, T.~N. Ferreira,
  M.~V.~M. Costa, and B.~Lee, ``A volumetric {SRP} with refinement step for
  sound source localization,'' \emph{IEEE Signal Processing Letters}, vol.~22,
  no.~8, pp. 1098--1102, 2015.

\bibitem{Nunes2012}
L.~O. Nunes, W.~A. Martins, M.~V.~S. Lima, L.~W.~P. Biscainho, B.~Lee, A.~Said,
  and R.~W. Schafer, ``Discriminability measure for microphone array source
  localization,'' in \emph{Proceedings of the International Workshop on
  Acoustic Signal Enhancement}, 2012, pp. 1--4.

\bibitem{Zhang2006}
L.~Zhang and X.~Wu, ``On the application of cross correlation function to
  subsample discrete time delay estimation,'' \emph{Digital Signal Processing},
  vol.~16, no.~6, pp. 682--694, 2006.

\bibitem{Lehmann2008}
E.~Lehmann and A.~Johansson, ``Prediction of energy decay in room impulse
  responses simulated with an image-source model,'' \emph{Journal of the
  Acoustical Society of America}, vol. 124, no.~1, pp. 269--277, 2008.

\end{thebibliography}

\end{document}